%% file: Main_Pap.tex
\documentclass[review]{elsarticle}
\usepackage[margin=1in]{geometry}
\usepackage{enumerate}
\usepackage{amsmath}
\usepackage{lineno,hyperref}
\usepackage{dingbat}
\usepackage{rotating}
\usepackage{graphicx}
\usepackage{pifont}
\usepackage{color}
\usepackage{rotating}
\modulolinenumbers[20]
\journal{arXiv}

\usepackage{array}
\usepackage{caption}
\newcolumntype{L}[1]{>{\raggedright\let\newline\\\arraybackslash\hspace{0pt}}m{#1}}

\begin{document}

\begin{frontmatter}

\title{Survey of prognostics methods for condition-based maintenance in engineering systems}

\author{Ehsan Taheri \corref{cor}\fnref{footnote1}}
\cortext[cor]{Corresponding author}
\fntext[footnote1]{Research fellow, Department of Aerospace Engineering, 1320 Beal Avenue, Ann Arbor, MI 48109, USA}
\ead{etaheri@umich.edu}

\author{Ilya Kolmanovsky \fnref{footnote2}}
\address{Department of Aerospace Engineering, University of Michigan, Ann Arbor, MI 48105, USA}
\fntext[footnote2]{Professor, Department of Aerospace Engineering, 1320 Beal Avenue, Ann Arbor, MI 48109, USA}
\ead{ilya@umich.edu}

\author{Oleg Gusikhin \fnref{footnote3}}
\address{Ford Research \& Advanced Engineering}
\fntext[footnote3]{Advanced Connected Services, Ford Research \& Advanced Engineering, Dearborn, MI, 48124, USA}
\ead{ogusikhi@ford.com}

\begin{abstract}
It is not surprising that the idea of efficient maintenance algorithms (originally motivated by strict emission regulations, and now driven by safety issues, logistics and customer satisfaction) has culminated in the so-called condition-based maintenance program. Condition-based program/monitoring consists of two major tasks, i.e., \textit{diagnostics} and \textit{prognostics} each of which has provided the impetus and technical challenges to the scientists and engineers in various fields of engineering. Prognostics deals with the prediction of the remaining useful life, future condition, or probability of reliable operation of an equipment based on the acquired condition monitoring data. This approach to modern maintenance practice promises to reduce the downtime, spares inventory, maintenance costs, and safety hazards. Given the significance of prognostics capabilities and the maturity of condition monitoring technology, there have been an increasing number of publications on machinery prognostics in the past few years. These publications cover a wide range of issues important to prognostics. Fortunately, improvement in computational resources technology has come to the aid of engineers by presenting more powerful onboard computational resources to make some aspects of these new problems tractable. In addition, it is possible to even leverage connected vehicle information through cloud-computing. Our goal is to provide a report on the state of the art and to summarize some of the recent advances in prognostics with the emphasis on models, algorithms and technologies used for data processing and decision making.
\end{abstract}


\end{frontmatter}

\linenumbers
\tableofcontents
\input{Introduction.tex}
\input{CBM-Models.tex}
\input{Application-Challenges.tex}

\input{Conclusion.tex}

\section*{References}

\bibliography{mybibfile}

\end{document}

%% file: Introduction.tex
\newpage
\section{Introduction}
Mechanical and electrical systems, and in particular, their building blocks/components, are subject to gradual tear and wear that will ultimately disrupt their proper operation and make them faulty. However, the deterioration procedure varies and depends on certain operating conditions such as stress, load and environment, etc. Considering the vast application and reliance of our daily life on machines, maintenance has a significant role in assuring safe and proper operation of the existing systems. 

Traditionally, maintenance activities have taken one of two approaches: preventive and corrective \cite{jardine2006review}. The (time- or duty-based) preventive maintenance also known as \textit{planned maintenance} defines a periodic time interval (or a certain duty), usually based on experience (or tests), to replace the component irrespective of its actual health status \cite{malik1979reliable}. For instance, the most common application of such a strategy, in automotive engineering, is the replacement of engine oil. These tasks are scheduled to occur after driving for a certain number of months (or miles). Another example is the timing belt on an automobile, which may be recommended to be replaced after five years (or 60,000 miles \cite{schwabacher2005survey}). 

Preventive maintenance leads to a costly maintenance strategy given the expenses associate with the modern complex components. In addition, the preventive maintenance does not provide any information about the health status of a component, which is a major defect for safety-critical components, which could lead to disasters, for instance, in the field of aerospace engineering. On the other hand, the corrective maintenance strategy seeks to replace a component once it is no longer operational and is not capable of performing its assigned task. This maintenance strategy, which is the most undesirable form of maintenance, has significant drawbacks. It is more labor intensive, does not eliminate catastrophic failures and causes unnecessary maintenance, which is costly by itself. In addition, there are costs associated with maintenance labor and downtime as well as the safety concerns and customer satisfaction. Considering a passenger vehicle, the impact on customer satisfaction is a major driving factor simply because the component failure might occur miles away from any repair shop. For other safety-critical applications (e.g. in the aerospace engineering), the corrective maintenance is avoided by adopting alternatives in which redundant components are considered since the failure is not tolerated. Collectively, expenses due to preventive and corrective maintenance constitute a significant portion of the expenses of many industrial companies.

\begin{figure}[htbp!]
\centering
\includegraphics[width=4.0in]{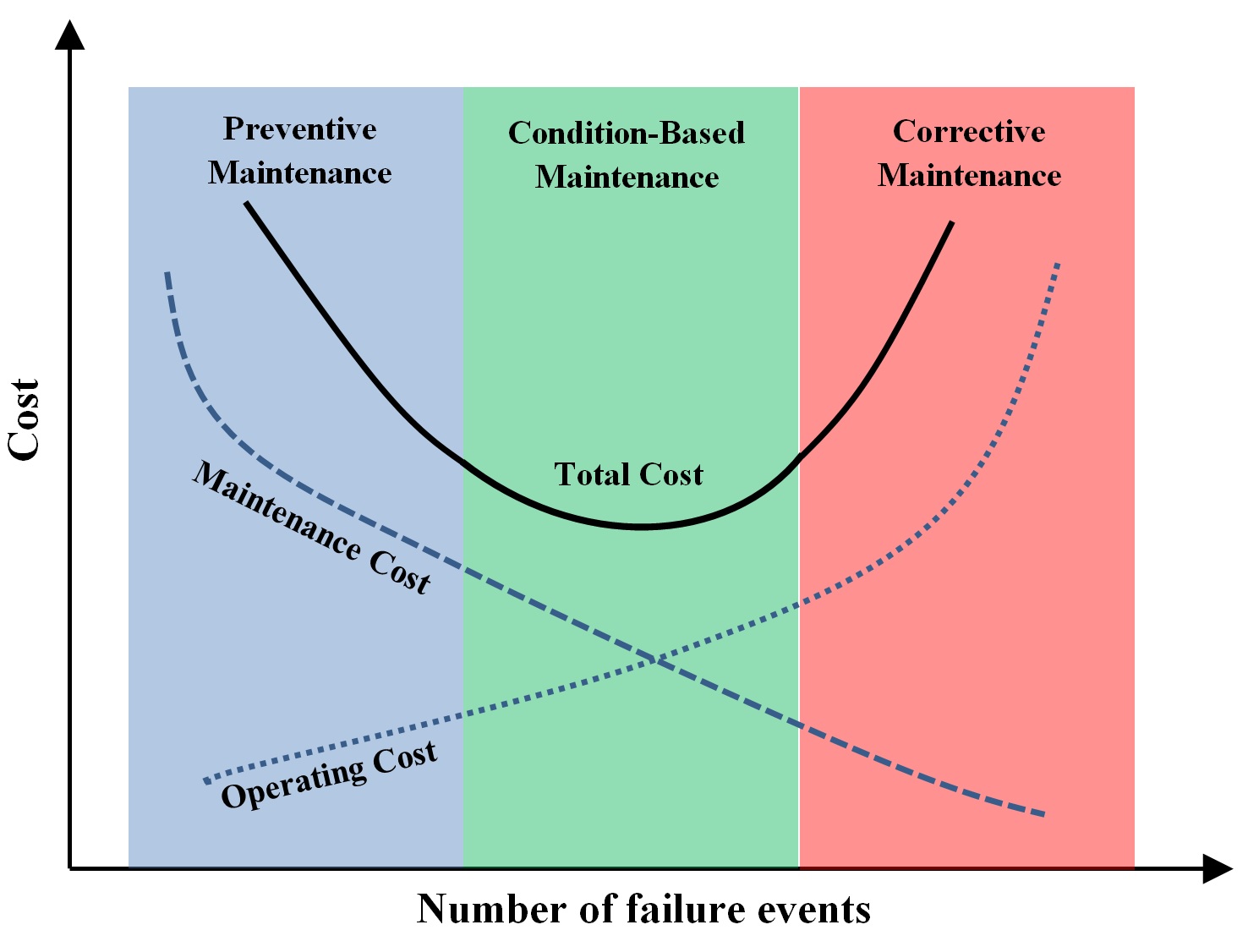}
\caption{Schematic diagram for operational, maintenance and total cost; colored regions denote different maintenance strategies.}
\label{fig:Maintenance}
\end{figure}

Between these two extreme maintenance strategies lies condition-based maintenance (CBM), wherein maintenance actions are performed as needed based on the condition of the equipment or component (see Fig.~\ref{fig:Maintenance}).
CBM avoids any unnecessary maintenance task by scheduling maintenance actions based on the conditions or observation of abnormal behaviours of a component. The more effective a CBM program is implemented the less maintenance cost will be. Within the aerospace community, and aside from the safety issue, this corresponds to lowering the downtime, which directly translates into significant amounts of money. With respect to the automotive industry, the replacement prices as well as the repair costs when multiplied by the population of vehicles is quite considerable. A CBM maintenance scheme can directly affect the following aspects of a system: 1) to improve the ability in detecting faults, 2) to improve the plant safety, 3) to better maintenance plans and decision making, 4) to reduce the inspection time and associated labor costs, 5) to increase the availability of assets.

CBM grants the ability to evaluate a system's actual health/damage conditions and provides the user with a prediction of failure, which is quite an essential tool for industrial applications. The costs associated to interruption of a business usually prove to be significantly higher than the expenses due to the repairs to return a business back to service \cite{orme2011property}. For the electrical machines the average annual rate of failure is estimated to be at least 3\% and for motors that have to operate under hostile conditions and environment, such as mining or pulp and paper industries, the annual failure rate is even greater and could be as high as 12\% \cite{sutherland2003prognostics}. Therefore, it is inevitable to ensure the availability of assets if a business is interested in profitable operation. This directly translates to an accurate estimation of the remaining useful life (RUL) of a system and its constituents or components. In other words, accurate RUL estimation can enable failure prevention in a more controllable manner in that effective maintenance can be executed in appropriate time to correct impending faults. There are two main tasks in a successful CBM, i.e., diagnostics and prognostics, which will be discussed in the next section. The overall life cycle cost of systems is reducible by implementing prognostics health monitoring (PHM) \cite{scanff2007life,feldman2009methodology}. On the other hand, developing a CBM is a significant technical challenge.

Presenting a survey for a field as diverse as CBM could be a daunting task. Perhaps the most difficult issue is restricting the scope of the survey to permit a meaningful discussion within a limited amount of space. To achieve this goal, we made a conscious decision to focus on the most important aspect of the CBM, i.e., prognostics. However, first we try to distinguish some of the salient aspects of diagnostics and its relation to prognostics. We, then, elaborate on the utmost objectives of CBM and highlight the significance of each task necessary for realizing any CBM program. A brief discussion of the methods, models and other important major steps of a CBM program are presented, whereas the emphasis is to provide the reader with a review that highlights and classifies the existing applications of prognostics in engineering areas such as aerospace, marine and automotive. We then discuss a few applications to prognostics of automotive engineering. Finally, we describe some of the challenges and opportunities that belong to the ongoing research. The authors' intent is to provide the researchers in scientific community, with the state-of-the-art in the aforementioned majors of engineering over the recent few years.


%% file: CBM-Models.tex
\section{Review of CBM, Modelings and Algorithms}\label{CBM}
In order to better understand the subject of CBM, it is necessary to distinguish between its two main constituents, i.e., \emph{diagnostics} and \emph{prognostics}. In the following sections we explain briefly the fundamental differences between these two tasks. We will also discuss the importance of the confidence limit, which is a major factor in the decision making procedure. Classification of the models used in prognostics is also provided.
\subsection{Diagnostics and Prognostics: Key Differences}
In principle, diagnostics is conducted to investigate the root cause of a failure and analyze the nature of a problem, whereas prognostics is related to predicting the future behaviour as a result of rational study and analysis of available pertinent data. Diagnostics itself is broken into three subtasks: 1) fault detection, 2) fault isolation, and 3) fault identification when it occurs \cite{jardine2006review}. Fault detection is a task to indicate whether something is going wrong in the monitored system; fault isolation deals with a task to locate the faulty component; and the last step, fault identification, is a task to determine the nature of the fault when it is detected. In terms of the relationship between diagnostics and prognostics, the former is an in-depth exploration of the failure mode to identify its leading cause after it has occurred within a system/component, whereas the latter is the process of generating a rational estimation of the RUL. Therefore, in its simplest form, prognostics is to monitor and detect the initial indications of degradation in a component, and be able to consistently make accurate predictions \cite{hess2002prognostics}. It is important to realize that time is a critical variable in prognostics and it is more or less trying to answer the question ``when a component will fail?'', distinguishing it from diagnostics, in which time plays a less important role, and instead the emphasis being placed more on determining the parameters of an already occurring fault or failure.

A diagnostics system consists of a series of steps each of which of its own importance. These steps include 1) data collection, 2) feature extraction (signal processing), and 3) a knowledge base of faults, which may be derived from expert knowledge, physical models and historical data. Therefore, it is highly reliant on the knowledge base as the final determination of what type of failure has occurred, and why it is achieved by comparing the utilizing feature extraction results with the knowledge base. A comprehensive review of techniques and methods used in fault diagnostics in beyond the scope of this work; however, the interested readers are referred to some of the excellent reviews \cite{jardine2006review,venkat2003reviewI, venkat2003reviewII,venkat2003reviewIII}.
The prognostics, on the other hand, shares some of the tasks of the diagnostics and requires several other steps. It shares the same tasks of feature extraction and a knowledge base of faults and further conducts performance assessment, degradation models, analysis of the degradation patterns and making judicious predictions. However, signals such as fault indicators and degradation rates, that the prognostics relies on, belong to the outputs of the diagnostics, which means that these two parts are somewhat intertwined. When combined, performance assessment and degradation models can describe a machine's relative health status and indicate what kind of degradation patterns may exist. The ultimate goal of most prognostic systems is accurate prediction of the RUL of individual systems or components, on the basis of their use and performance. This is important since it allows advances scheduling of maintenance activities, proactive allocation of replacement parts and enhances fleet deployment decision based on the estimated progression of component life. Prediction algorithms, which could be derived from classic time series theories, statistics or artificial intelligence technologies, can forecast when machine performance will decrease to an unacceptable level as defined by the failure analysis and health management.

Engineering prognostics is used by industry to reduce business risks due to unexpected failures of equipment. It still relies highly on the experience and knowledge gained over years and its application is limited to systems for which significant data base is available (e.g., rotary machines). On the other hand, the models used in prognostics are application dependent, which requires extensive analysis of the results and assumptions. Appropriate model selection for successful practical implementation, requires both a mathematical understanding of each model type, and also an appreciation of how a particular business intends to utilize the models and their outputs. Unfortunately, there is no general prognostic model to fit all business needs and not all of the models are well proven mathematically. In addition, efficacy of models is dependent upon the availability of required data, skilled personnel and computing infrastructure.

Prognostics is a relatively new research area and is not a well-developed discipline compared to other areas of CBM. A number of literature reviews covering CBM with emphasis on prognostic components including models and approaches have already been presented in \cite{scarf1997application,monitoring1diagnostics, pusey1999assessment, engel2000prognostics,luo2003interacting, heng2009rotating, vachtsevanos13intelligent, sun2012benefits, lee2014prognostics}. Table \ref{tab:litreview} summarizes some of the most important review papers to date where AI, SA and ANN stand for Artificial Intelligence, Signal Analysis and Artificial Neural Networks, respectively. In addition, Reference \cite{si2011remaining} reviews the benefits and challenges of prognostics and Reference \cite{prajapati2012condition} reviews the condition-based maintenance. Table \ref{tab:PredictiveTechnologies} also shows highlights typical applications for some of the more common predictive maintenance technologies \cite{maintenance2008guide}.

\begin{table}[htbp!]
\caption{Summary of the existing review papers and their focus on different prognostics methods.}
\label{tab:litreview}
\begin{tabular}{l l l l l l l l}
\hline
\hline
Reference                     & Year & Knowledge- & Experience- &  Data-      & Model-     & Hybrid     & Other methods \\
                              &      & based      &    based    &  Driven     & based      &            &               \\
\hline
\cite{luo2003model} & 2003    &      &            &  \checkmark & \checkmark  &            &       \\
\cite{schwabacher2005survey}  & 2005 &            &             &  \checkmark &            &            &       \\
\cite{jardine2006review}      & 2006 &            &             &  \checkmark & \checkmark &            & AI \\
\cite{lee2006intelligent}     & 2006 &            &             &  \checkmark & \checkmark & \checkmark &  \\
\cite{goh2006review}          & 2006 &            &             &  \checkmark & \checkmark &            &  \\
\cite{kothamasu2009system}    & 2006 &            &             &             & \checkmark &            &  Reliability, Stochastic\\
\cite{coble2008prognostic}    & 2008 &            &             &  \checkmark &            &            &  Stress and effects-based \\
\cite{heng2009rotating}       & 2009 &            &             &  \checkmark & \checkmark & \checkmark &  \\
\cite{sikorska2011prognostic} & 2011 & \checkmark &             &             & \checkmark &            & Life Expectancy, ANN      \\
\cite{si2011remaining}        & 2011 &            &             &  \checkmark &            &            &  \\
\cite{ahmadzadeh2014remaining}& 2014 & \checkmark & \checkmark  &  \checkmark & \checkmark & \checkmark &  \\
\cite{tsui2015prognostics}    & 2014 &            &             &  \checkmark & \checkmark &            &  SA, Stochastic, ANN\\
\cite{tsui2015prognostics}    & 2015 &            &             &  \checkmark &            &            &  \\

\hline
\end{tabular}
\end{table}

\begin{table}[htbp!]
\centering
\caption{Common predictive technology applications.}
\label{tab:PredictiveTechnologies}
\begin{tabular}{l l l l l l l l l l l l }
\hline
\hline
Technology   & \rotatebox{90}{Pumps} & \rotatebox{90}{Electric Motors} & \rotatebox{90}{Diesel Generators} & \rotatebox{90}{Condensers} & \rotatebox{90}{Heavy Equipment/Cranes} & \rotatebox{90}{Circuit Breakers} & \rotatebox{90}{Valves} & \rotatebox{90}{Heat Exchangers} & \rotatebox{90}{Electrical Systems} & \rotatebox{90}{Transformers} & \rotatebox{90}{Tanks, Piping} \\
\hline
Vibration Monitoring/Analysis & $\times$ & $\times$ & $\times$ & & $\times$ & & & & & & \\
Lubricant, Fuel Analysis      & $\times$ & $\times$ & $\times$ & & $\times$ & & & & &$\times$ & \\
Wear Particle Analysis        & $\times$ & $\times$ & $\times$ & & $\times$ & & & & & & \\
Bearing, Temperature/Analysis & $\times$ & $\times$ & $\times$ & & $\times$ & & & & & & \\
Performance Monitoring        & $\times$ & $\times$ & $\times$ & $\times$ &  & & & $\times$ & & $\times$ & \\
Ultrasonic Noise Detection    & $\times$ & $\times$ & $\times$ & $\times$ &  & & $\times$ & $\times$ & & $\times$ & \\
Ultrasonic Flow               & $\times$ &  &  & $\times$ &  & & $\times$ & $\times$ & &  & \\
Infrared Thermography         & $\times$ & $\times$ & $\times$ & $\times$ & $\times$ & $\times$ & $\times$ & $\times$ & $\times$ & $\times$ & \\
Non-destructive Testing (Thickness) &  & & & $\times$ &  & &  & $\times$ & &  & $\times$\\
Visual Inspection             & $\times$ & $\times$ & $\times$ & $\times$ & $\times$ & $\times$ & $\times$ & $\times$ & $\times$ & $\times$ & $\times$ \\
Insulation Resistance         &   & $\times$ & $\times$ &   &   & $\times$ &   &   & $\times$ & $\times$ &   \\
Motor Current Signature Analysis &   & $\times$ &   &   &   &   &   &   &   &   &   \\
Motor Circuit Analysis        & & $\times$ &   &   &   & $\times$  &   &   &  $\times$ &   &   \\
Polarization Index            & & $\times$ & $\times$  &   &   &   &   &   &  $\times$ &   &   \\
Motor Circuit Analysis        & &  &   &   &   &   &   &   &  $\times$ & $\times$  &   \\
\hline
\end{tabular}
\end{table}
Although useful in appreciating the state of the art, we feel that there is a need for a literature review that incorporated the salient aspects of a reliable CBM that not only presents a review of models and their merits but also focuses on specific practical implementations in specific engineering fields.
Reference \cite{lee2014prognostics} adopts a similar strategy while focusing on rotary machine systems whereas the application of prognostics for other components is growing. Knowledge of the prior work is a necessity for future research efforts. To address this gap, this paper provides a review of the field of PHM, which focuses on the practical applications on various components in the fields of engineering such as automotive, aerospace and marine engineering.
\subsection{Critical component identification}
\begin{figure}[htbp!]
\centering
\includegraphics[width=4.0in]{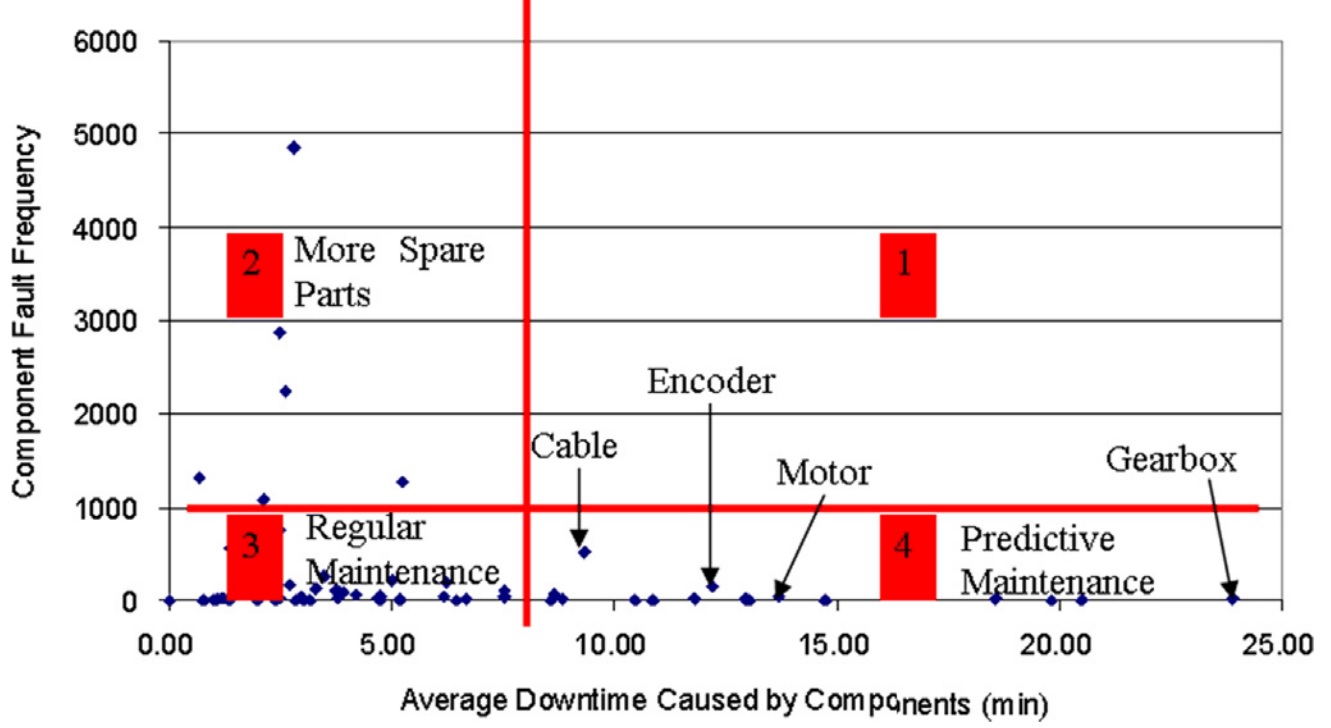}
\caption{Component fault Frequency-Downtime chart and four quadrants for identifying critical components \cite{lee2014prognostics}.}
\label{fig:FourQuadrantChart}
\end{figure}

Identifying critical components is the first step in developing a prognostics and health monitoring system. One approach in identifying the significance of components on the overall performance and cost downtime of a system is to use a quadrant chart as is shown in Fig.~\ref{fig:FourQuadrantChart} (taken from \cite{lee2014prognostics}). A similar figure is also shown in \cite{brahimi2016critical} for the selection of critical components.
It displays the frequency of failure versus the average downtime associated with failure for relevant components of 890 SW Robots. The effectiveness of the current maintenance strategy can be seen when the data is graphed in this way. The horizontal and vertical lines
that divide the graph to four quadrants are user-defined parameters based on their demands on production and/or maintenance. The resulting quadrants are numbered 1-4 starting with the upper right and moving counter clockwise. The first quadrant represents those components that not only fail more frequently, but also result in extensive downtime. Typically, there should not be any components in this quadrant because such issues should have been noticed and fixed during the design stage. However, there could be instances in which a manufacturing defect in, or continued improper use of, a particular component could result in repetitive failures and significant downtime. The second quadrant still contains components with a high frequency of failure, but each component causes a short downtime. The maintenance recommendation for such components is to have an adequate number of spare parts on hand. The third quadrant contains components with a low frequency of failure and low average downtime per failure, which means that the current maintenance practices are working for these components and no changes are required. In the fourth quadrant lie the most critical components as their
failures, though infrequent, cause the most downtime per occurrence and could potentially incur significant costs. The components of this last quadrant should be the focus of prognostics. For instance, as is shown in Fig.~\ref{fig:FourQuadrantChart}, the filed of robotics prognostics should focus on encoder, motor and gearbox as critical components. The existence of similar data for the other engineering fields improves the return of prognostics by developing frameworks for components that play a critical role in the overall performance and cost. The reader is referred to \cite{lee2014prognostics} for additional information on this matter.

\subsection{Failure modes and Prognostic tasks}
\begin{figure}[htbp!]
\centering
\includegraphics[width=5.0in]{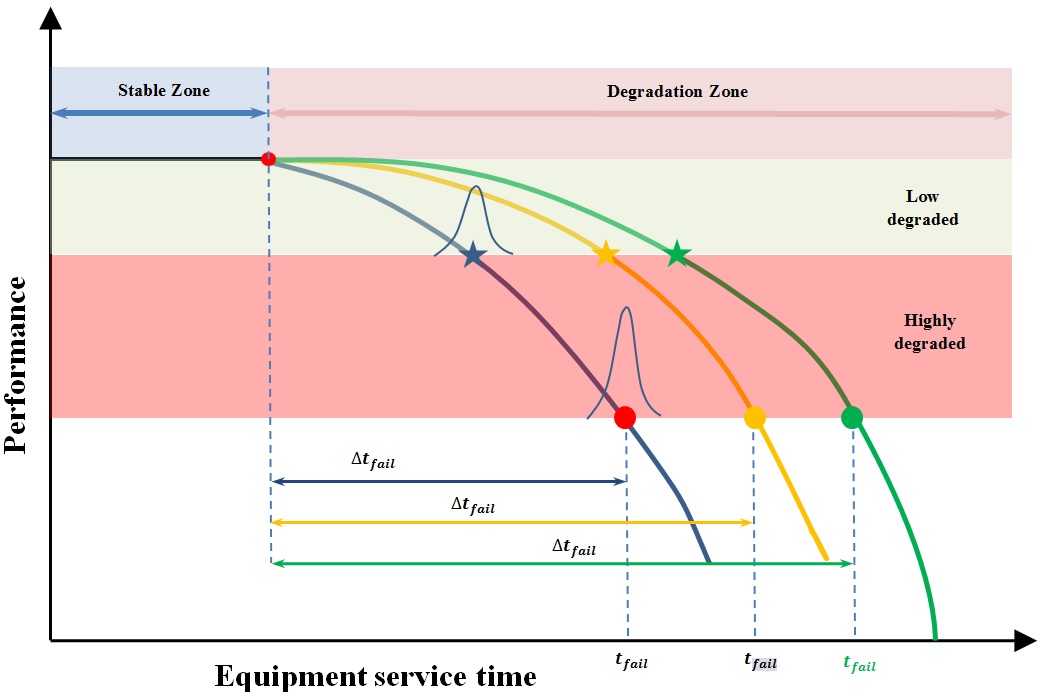}
\caption{Typical performance degradation for three different failure modes.}
\label{fig:FailureModes}
\end{figure}

To understand the role of models in prognostics, it is important to identify the various steps involved in obtaining an RUL estimate (which is the holy grail of prognostics) and its confidence bounds. Figure.~\ref{fig:FailureModes} shows the process a component undergoes from a healthy state performance until its final failure.
It depicts highly simplified degradation curves for three different and independent failure modes, which could represent different types of failure of the same component. There is a stable zone during which the performance of the component is not affected. However, the component is eventually going to degrade and fall into the degradation zone, which itself is divided into two regions, i.e.,- low and high-degraded regions defined by their bounds (levels), respectively. The selection of the performance levels is a critical task in any prognostics approach. In addition, Fig.~\ref{fig:FailureModes} shows the spread of the time once a degradation curve hits a specific level. The confidence (precision) in determining those probabilities is essential in decision making and is discussed in the next section. Note also that there are factors that effect the degradation patterns which triggers different failures. The progression of any failure mode may be accelerated due to the changes in the operating conditions, maintenance actions or even progression of the other failure modes (e.g, a bearing fault causes high vibration that induces and accelerates mechanical seal degradation). Therefore, an efficient procedure to estimate RUL correctly needs to address the following questions (or know preliminary information about them): 1) what is the current degradation rate?, 
2) which failure mode (or modes) has (have) been triggered and contributes to the degradation?, and 3) 
how much is known of the severity of the degradation? (determines the position of the component of the particular curve).

If a systems-oriented approach to prognostic-based decision support is desired, then RUL estimates should be further supplemented with forecasts describing the impact of predicted failures on operational and maintenance activities which can be considered at the business management level rather than prognostics task \cite{khalak2006influence,julka2011making}. Based on the collective approaches, one could conceptualize a diagnostic/prognostic framework that addresses prognostics through three levels with varying degrees of complexity, i.e., existing failure mode prognostics, future failure mode prognostics and post-action prognostics \cite{sikorska2011prognostic}. The prognostics models discussed in this review keeps the complexity to the simplest level, i.e., existing failure mode prognostics. Almost all of the works in the literature belong to this category.

\subsection{Confidence limits}
The output of a prognostic algorithm has two components: 1) an estimate of time to failure, which is also referred to as the RUL and 2) an associated confidence limit \cite{engel2000prognostics}. Analysis of the confidence limit is important since the prognostics intrinsically deals with estimating an uncertain variable parameter, which is effected by several factors including the future operation of the component, operating conditions and errors due to the fidelity of the utilized diagnostics and prognostics models. Confidence limits are even more important in prognostic modelling than for diagnostic prediction. This is due to the fact that in diagnostics the failure and the extent of damage is known and is an externally verifiable quantity (e.g., actual crack size) whereas this is not the case in prognostics as it deals with failure. It is highly important for business decisions to be made based on the bounds of the RUL confidence interval rather than a specific value of the component expected life \cite{engel2000prognostics}.

\subsection{Implementing prognostic models}
There are certain aspects that have to be considered before implementing any prognostics model. First of all, most of the prognostics program deal with accurate prediction of URL of an identified failure mode. This strategy is retained to keep the process simple and tractable. In addition, the existence of certain type of data, level of complexity of the model and the underlying assumptions will make the models better suited to certain applications. One could pose a series of questions to assess the performance and suitability of a particular model to a particular problem,
\begin{itemize}
\item Prediction requirement: what does the RUL prediction need to achieve?
\item Model-process capability: can the model describe the reality?
\item Resource requirements: are the resources available to undertake the modelling?
\item Approach readiness: is the modelling approach sufficiently well proven to be relied upon?
\end{itemize}
These four criteria do not include factor that should be considered before prognostics are undertaken in the first place, which is beyond the scope of this work. A good discussion of this topic is presented in \cite{carnero2006evaluation}.

\section{Prognostic Models And Their Classification}
\begin{figure}[htbp!]
\centering
\includegraphics[width=6.0in]{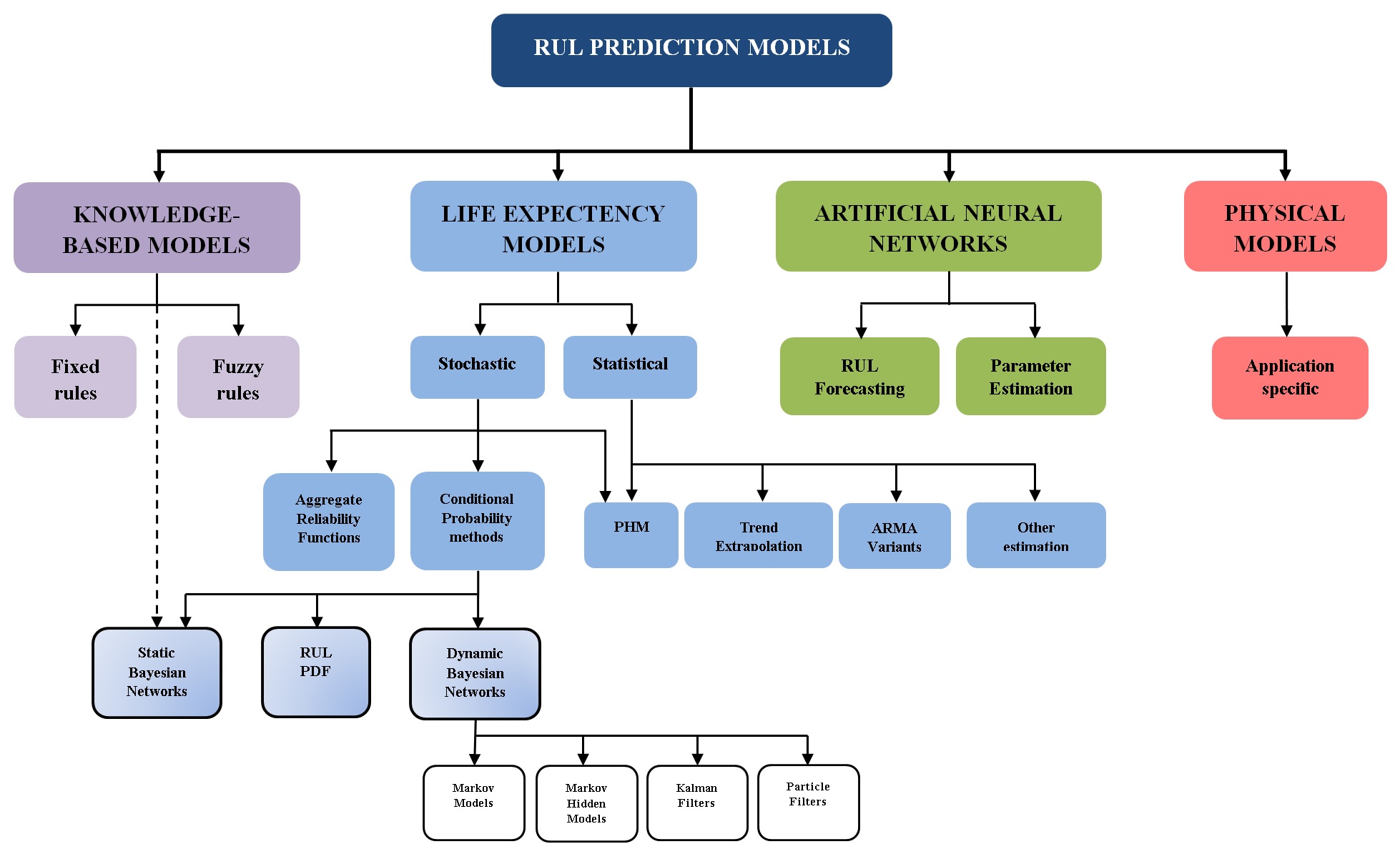}
\caption{Classification of models used for URL prediction.}
\label{fig:URLModels}
\end{figure}

With the discussion given on the overall task of prognostics and the importance of the RUL estimation we focus on the existing models and their capability in providing the necessary information to practitioners. Current prognostic approaches can be categorized into four major classes: experimental, data-driven, model-based and hybrid. Reviewing the literature it is apparent that papers limit their discussions to data-riven or model-based approaches and a few of them address the experience-based approach. A detailed classification of models into four groups is given in \cite{sikorska2011prognostic} specifically designed for RUL prediction. It is further divided into varying number of subgroups (see Fig. \ref{fig:URLModels}). The material of this section is mainly taken from \cite{sikorska2011prognostic}.
\begin{itemize}
\item Knowledge-based models \cite{biagetti2004automatic,slottner2008knowledge,gorodetskii1997generating, biagetti2004automatic, abele2013knowledge}: these models assess the similarity between an observed situation and a database of previously defined failures and deduce the life expectancy from previous events. Sub-categories include \emph{expert systems} and \emph{Fuzzy systems}.

\item Life expectancy models: determine the life expectancy of individual machine components with respect to the expected risk of deterioration under known operating conditions. Sub-categories are separated into \emph{statistical} and \emph{stochastic} models. Stochastic models are further divided into two models i.e., aggregate reliability functions and conditional probability methods. Statistical models include trend extrapolation, auto-regressive moving average (ARMA) model and its variants, and proportional hazard modelling (PHM).

\item Artificial Neural Networks: These models compute an estimated output for the RUL of a component, directly or indirectly, from a mathematical representation of the component that has been derived from observation data rather than a physical understanding of the failure processes. They are further grouped into models used for direct URL forecasting and parameter estimation for other models.

\item Physical models: These models compute an estimated output for the RUL of a component from a mathematical representation of the physical behavior of the degradation process. Types of physical models tend to be application (failure mode) specific and are therefore not classified further.
\end{itemize}
It is a difficult task to strictly categorize a model into the presented classes, particularly due to the fact that more recently the models in the literature are a combination of two or more classical modelling approaches.
Model selection requires that the main advantages and disadvantages of each model type be well understood. For a list of generic advantages and disadvantages of the introduced models refer to \cite{sikorska2011prognostic}. A brief description of each model is given in Table~\ref{tab:ModelsAdvantages} to familiarize the reader with their basic advantages and disadvantages. The advantages and disadvantages are mostly associated with the simplicity of the method (either model and/or its of implementation), capability to provide confidence limit, reliance on the amount and accuracy of data, availability of tools and softwares, capability to incorporate new data, ability to model previously
unanticipated faults, capability to manage incomplete data sets, being able to model multivariate dynamic models and their level of computational efficacy. Table \ref{tab:whentouse} summarizes considerations for using or avoiding a particular type of model. These two tables serve as an initial guideline for selecting a particular model to be used in the later stages of a prognostic framework. In the next sections we briefly discuss the underlying principles of each model.
\begin{table}[htbp!] %
\centering
{\tiny
\caption{Advantages and disadvantages of prognostic modelling options.}
\label{tab:ModelsAdvantages}
\begin{tabular}{L{2.0cm} L{4.5cm} L{4.5cm}}
\hline
\hline
Model & Advantages & Disadvantages \\
\hline
\textbf{Knowledge based} & & \\
Expert systems & \begin{itemize} \item Simple (albeit time consuming) to develop \item  Easy to understand \end{itemize} & \begin{itemize} \item Relies entirely on knowledge of subject matter experts \item Significant number of rules required \item Significant management overhead to keep knowledge base up to date \item Precise inputs required \item No confidence limits supplied  \item Not feasible to provide exact RUL output \end{itemize}\\
Fuzzy systems & \begin{itemize} \item Fewer rules required than for expert systems \item Inputs can be imprecise, noisy or incomplete \item confidence limits can be provided on the output with some types of models \end{itemize} & Domain experts required to develop rules \\
\end{tabular}
}
\end{table}

\begin{table}[htbp!] %
\centering
{\tiny
\caption*{Table \ref{tab:ModelsAdvantages} (continued).}
\begin{tabular}{L{2.0cm} L{4.5cm} L{4.5cm}}
\hline
\hline
Model & Advantages & Disadvantages \\
\hline
\textbf{Stochastic} & & \\
Aggregate reliability functions & \begin{itemize} \item Simple and well understood by reliability engineering community \item Numerous software options available \item Theoretically can be performed at all equipment hierarchy levels, especially whin a small number of failure modes dominate \item Confidence limits are available for RUL predictions \item Accuracy and precision increases as RUL decreases resulting in the ability to set useful warning limits \end{itemize} & \begin{itemize} \item Failures must be statistically independent and identically distributed \item In most cases will require a statistically significant sample size pertaining to each failure mode for reliable RUL predictions \item Warnings prior to actual failure are not readily available \end{itemize}\\
\hline
RUL PDF & \begin{itemize} \item Simple and easy adaptation of basic reliability approaches \item Only requires that time at which failure has not occurred is monitored (i.e, no condition monitoring data) \item Theoretically can be performed at all equipment hierarchy levels, especially when a small number of failure modes dominate \item Confidence limits are available for RUL predictions \item Accuracy and precision increases as RUL decreases resulting in the ability to set useful warning limits \end{itemize} & \begin{itemize} \item Available accuracy and precision is dependent on forecasting interval  \item In most cases will require a statistically significant sample size pertaining to each failure mode for reliable RUL predictions \item Assumes that hazard is a function of operating time rather than external risk factors \end{itemize} \\
\end{tabular}
}
\end{table}

\begin{table}[htbp!] %
\centering
{\tiny
\caption*{Table \ref{tab:ModelsAdvantages} (continued).}
\begin{tabular}{L{2.0cm} L{4.5cm} L{4.5cm}}
\hline
\hline
Model & Advantages & Disadvantages \\
\hline
\textbf{Stochastic} & & \\
Static Bayesian Networks & \begin{itemize} \item Can readily manage incomplete data sets \item Allow/force user to learn about causal relationships \item Captures and integrates expert knowledge \item Algorithms available to avoid the over fitting of data \item Computer software available for modelling \item Confidence limits are intrinsically provided \end{itemize} & \begin{itemize} \item Cannot model previously unanticipated faults and/or root causes \item Computational difficulty of exploring a previously unknown network \item A Bayesian network is only as useful as the prior knowledge is reliable \item Results may be sensitive to selection of prior distribution \item Modelling experts required in addition to domain experts \end{itemize}\\
\hline
Markov, Semi-Markov models & \begin{itemize} \item Well established approach and able to model numerous system designs and failure scenarios \item Can readily manage incomplete data sets \end{itemize} & \begin{itemize} \item Reasonably large volume of data required for training \item Assumes a single monotonic, non temporal failure degradation pattern (i.e., different stages of failure cannot be accounted for) \item Cannot model previously unanticipated faults and/or root causes \item More complex semi-Markov models are required if failures or failur progression times are not exponentially distributed \item Not appropriate for repairable systems that are only partially restored \end{itemize}\\
\end{tabular}
}
\end{table}

\begin{table}[htbp!] %
\centering
{\tiny
\caption*{Table \ref{tab:ModelsAdvantages} (continued).}
\begin{tabular}{L{2.0cm} L{4.5cm} L{4.5cm}}
\hline
\hline
Model & Advantages & Disadvantages \\
\hline
\textbf{Stochastic} & & \\
Hidden Markov, Semi-Markov models & \begin{itemize} \item Can model different stages of degradation so failure trend does not need to be monotonic \item Can model spatial and temporal data \item Specific knowledge of failure mechanism progression is not required \item Can readily manage incomplete data sets \item Provide confidence limits as part of their RUL prediction \end{itemize} & \begin{itemize} \item Large volume of data required for training, proportional to the number of hidden states \item Cannot model previously unanticipated faults and/or root causes \item More complex Hidden semi-Markov models are required if failures or failure progression times are not exponentially distributed \item Computationally intensive, particularly for a large number of hidden states \end{itemize}\\
Bayesian techniques with Kalman Filters & \begin{itemize} \item Can be used to model multivariate, dynamic processes \item Basic KF is computationally efficient, particulary for systems with a large number of states \item Can accommodate incomplete and noisy measurements \item Variants available for non-linear processes \item Other advantages on underlying Bayesian technique \end{itemize} & \begin{itemize} \item Process and measurement noise must be Gaussian \item Some variants diverge easily \item Variants for non-linear systems are more computationally intensive than basic Kalman filters \item Measurement data required \item Other disadvantages depend on underlying Bayesian technique \end{itemize}\\

\end{tabular}
}
\end{table}

\begin{table}[htbp!] %
\centering
{\tiny
\caption*{Table \ref{tab:ModelsAdvantages} (continued).}
\begin{tabular}{L{2.0cm} L{4.0cm} L{5cm}}
\hline
\hline
Model & Advantages & Disadvantages \\
\hline
\textbf{Stochastic} & & \\
Bayesian techniques with Particle Filters & \begin{itemize} \item Can bes used to model multivariate, dynamic processes \item Noise does not need to be either linear of Gaussian \item More accurate than Kalman filter variants for non-linear systems \item Other advantages depend on underlying Bayesian technique \end{itemize} & \begin{itemize} \item A large number of samples (or resampling) are required to avoid degeneracy problem \item Can be more computationally intensive than basic Kalman filters \item Measurement data required \item Other disadvantages depend on the underlying Bayesian technique  \end{itemize}\\
\hline
\textbf{Statistical} & & \\
Trend extrapolation & \begin{itemize} \item Simplest technique to apply and explain \item Easy to set alarms \item Advanced software tools not required \end{itemize} & \begin{itemize} \item Few failures have a well-defined monotonic, single-parameter trend \item Interpretability is affected by process/measurement noise and variations in operating conditions \item Availability of confidence limits dependent on amount of data at the different states of failure development \end{itemize}\\
ARMA Models \& variants & \begin{itemize} \item Advanced ARMA related techniques available for non-stationary data \item Historical failure data is not required \item Usually computationally efficient and therefore can be performed in real time \item An understanding of detailed failure mechanisms not required \item Provide accurate and reliable short term  predictions of RUL \end{itemize} & \begin{itemize} \item Basic ARMA models assume stationarity of the process and noise \item Does not integrate prior or expert knowledge \item sensitive to noise and initial conditions \item significant data required for model development and validation \item Long-term predictions of RUL are less reliable \end{itemize} 
\end{tabular}
}
\end{table}

\begin{table}[htbp!] %
\centering
{\tiny
\caption*{Table \ref{tab:ModelsAdvantages} (continued).}
\begin{tabular}{L{2.0cm} L{4.0cm} L{5cm}}
\hline
\hline
Model & Advantages & Disadvantages \\
\hline
\textbf{Statistical} & & \\
PHM & \begin{itemize} \item COTS software available \item Accounts for age dependent and independent hazards \item Models are simple to develop Confidence limits can be calculated \end{itemize} & \begin{itemize} \item All relevant covariates must be included in the model \item Mixing different types of covariates in one model may be problematic  \item Strict (albeit implied) assumptions regarding nature of underlying process \item Historical data required pertaining to individual failure modes \item Multi-collinearity, monotonicity and large covariate values that can cause a failure of the model parameter estimation process \item Parameter selection often manual and time consuming and the selection of parametric estimation technique is not straightforward \item Traditional PHM equation assumes covariates describe a stationary process. Dynamic PHM is more involved \item Can only be used to develop models for failures that have been experienced previously and for which associate covariate data is available \item Too easy to develop a model that may be statistically adequate but does not represent any actual failure phenomenon (i.e. physically meaningless) \end{itemize}\\
\end{tabular}
}
\end{table}

\begin{table}[htbp!] %
\centering
{\tiny
\caption*{Table \ref{tab:ModelsAdvantages} (continued).}
\begin{tabular}{L{2.0cm} L{4.0cm} L{5cm}}
\hline
\hline
Model & Advantages & Disadvantages \\
\hline
\textbf{Artificial Neural Networks} & &\\
For Forecasting with ANNs & \begin{itemize} \item Complex multi-dimensional, non-linear systems can be modelled \item Physical understanding of the system behaviour not required \item ANN variants facilitate the use of any type of input data \item Computer software is available for modelling \end{itemize} & \begin{itemize} \item Requires a significant amount of data for training data that needs to be representative of true data range and its variability \item Determining the most appropriate model is  largely trial and error and therefore can be time consuming \item Most networks cannot provide confidence limits on the output \item Pre-processing is required to limit the number of data inputs and reduce model complexity \item All published research is relatively recent \item Outputs need to mapped to a physical representation \end{itemize}\\
Parameter Estimation with ANNs & \begin{itemize} \item As for RUL Forecasting with ANNs \item Useful for incorporating with physics of failure models \item Confidence limits available from underlying model (for which parameters are being estimated) \end{itemize} & \begin{itemize} \item Less data required for estimating parameters as models tend to be failure specific \item Determining the most appropriate model is largely trial and error and therefore can be time consuming \end{itemize}\\
\hline
\textbf{Physical models} &&\\
Physical Models & \begin{itemize} \item Provide most accurate and precise estimates of all  modelling options \item Confidence limits provided \item Outputs can be easily understood \end{itemize} & \begin{itemize} \item Detailed and complete knowledge of system behaviour required \item The accuracy and robustness are subject to the experimental conditions under which models were developed \end{itemize} \\
\end{tabular}
}
\end{table}

\begin{table}[htbp!] %
\centering
{\tiny
\caption{When to consider/avoid using particular models.}
\label{tab:whentouse}
\begin{tabular}{L{2.0cm} L{4.5cm} L{4.5cm}}
\hline
\hline
Model & When to consider & When to avoid \\
\hline
\textbf{Knowledge based} & & \\
Expert systems & \begin{itemize} \item Well-understood, stable, narrow problem area \item human experts are available to develop the knowledge base; and operating conditions are stable and predictable; and simple precise queries to define potential faults is impossible; and only an approximate RUL estimate is required \end{itemize} & \begin{itemize} \item No human experts are available to define comprehensive set of rules; or fault maintenance are not well understood; or operating conditions are highly variable; or highly accurate or precise RUL estimates are required \end{itemize}\\
Fuzzy systems & \begin{itemize} \item One or more variables are continuous; and a mathematical model is not available or not feasible to implement; and data contains high levels of noise or uncertainty; and difficult to define exact queries that identify specific faults \end{itemize} & No human experts are available to define fuzzy rules; or input data is discrete and limited to a small number of options \\
\hline
\textbf{Stochastic} & & \\
Aggregate reliability functions & \begin{itemize} \item Sample size is statistically significant and representative of individual sample; and \item Small set of dominant failure modes; and \item PDF is not exponential; and Reliability growth is not occurring; and \item RUL prediction is predominantly used for overall maintenance management rather than tracking of a specific asset (e.g., when redundancy is available) so gradual escalation of warning levels are not required \end{itemize} & \begin{itemize} \item Only a small number of failures can be attributed to individual failure modes; or \item Significant number of possible failure modes that cannot be easily differentiated, or historically have not been; or \item Hazard rate is constant; or \item Past operating conditions are not representative of current environment or usage; or \item The specific asset is critical to plant safety or operations and warning is required prior to failure \end{itemize}\\
\end{tabular}
}
\end{table}
\begin{table}[htbp!] %
\centering
{\tiny
\caption*{Table \ref{tab:whentouse} (continued).}
\begin{tabular}{L{2.0cm} L{4.5cm} L{4.5cm}}
\hline
\hline
Model & When to consider & When to avoid \\
\hline
\textbf{Stochastic} & & \\
RUL PDF & \begin{itemize} \item Sample size is statistically significant and representative of individual sample; and Small set of dominant failure modes; and \item PDF is not exponential; and \item Reliability growth is not occurring \item Condition monitoring data is not available; and \item Operating age can be tracked to confirm absence of failure; and \item Only final estimates need to be particularly accurate and precise \end{itemize} & \begin{itemize} \item Only a small number of failures can be attributed to individual failure modes; or \item Significant number of possible failure modes that cannot be easily differentiated, or historically have not been; or \item Hazard rate is constant; or \item Past operating conditions are not representative of current environment or usage; or \item Failure is hidden and no failure finding is being undertaken; or \item High-level of accuracy and precision is required a long time into the future \end{itemize} \\
\end{tabular}
}
\end{table}

\begin{table}[htbp!] %
\centering
{\tiny
\caption*{Table \ref{tab:whentouse} (continued).}
\begin{tabular}{L{2.0cm} L{4.5cm} L{4.5cm}}
\hline
\hline
Model & Advantages & Disadvantages \\
\hline
\textbf{Stochastic} & & \\
Static Bayesian Networks & \begin{itemize} \item Incomplete, multivariate data available; and \item Root cause of failure known; and \item process and plant configuration is relatively static or network is confirmed up to date; and \item Modelling experts are available \end{itemize} & \begin{itemize} \item Root causes of failure unknown; or \item Expert plant and modelling knowledge unavailable; or \item Training data is unavailable \end{itemize}\\
\hline
Markov, Semi-Markov models & \begin{itemize} \item Simple to develop and implement; \item Incomplete, multivariate data available; and Root causes of failure known; and \item Process and plant configuration is relatively static or network is confirmed up to date; and \item Relatively accurate and precise RUL estimate is required \end{itemize} & \begin{itemize} \item Repairable system; or \item Temporal measurement data as model inputs; or \item Sufficient data related to failure mode is not available for training; or \item Failure being modelled has more than one discrete stage (e.g., crack initiation, growth , final failure, etc) \end{itemize}\\
\hline
Hidden Markov, Semi-Markov models & \begin{itemize} \item Repairable systems; and \item Root causes of failure known; and \item Failure being modelled has more than one discrete stage \item Temporal data to be used as model inputs \item Relatively accurate and precise RUL required \end{itemize} & \begin{itemize} \item Sufficient data related to failure mode is not available for training; or \item Suitable hardware for computation is not available \end{itemize}\\
\hline
Bayesian techniques with Kalman Filters & \begin{itemize} \item Multivariate posterior distribution; and \item Additive; and \item Condition monitoring data is available; and \item Relatively accurate and precise RUL estimate required \end{itemize} & \begin{itemize} \item Multiplicative noise; or \item Single variable posterior distribution; or \item Covariate data is not available for the failures of interest \end{itemize}\\

\end{tabular}
}
\end{table}

\begin{table}[htbp!] %
\centering
{\tiny
\caption*{Table \ref{tab:whentouse} (continued).}
\begin{tabular}{L{2.0cm} L{4.0cm} L{5cm}}
\hline
\hline
Model & Advantages & Disadvantages \\
\hline
\textbf{Stochastic} & & \\
Bayesian techniques with Particle Filters & \begin{itemize} \item Multi-variate or non-standard posterior distribution \item Non-linear, non-Gaussian noise; and \item Relatively accurate and precise RUL estimate required \end{itemize} & \begin{itemize} \item Typical deterministic posterior distribution; or \item Linear, Gaussian; or \item Multiplicative noise; or \item Single variable posterior distribution; or \item Covariate data is not available for the failures of interest \end{itemize}\\
\hline
\textbf{Statistical} & & \\
Trend extrapolation & \begin{itemize} \item Single defined failure mode associated with a single monitored (or calculated) parameter that can be described with a monotonic trend; and operating conditions are stable or do not affect monitored parameter; and \item Measurements are repeatable, reliable and not highly sensitive to measurement processes (e.g., online sensors) \end{itemize} & \begin{itemize} \item Incipient failure cannot be related to a simple measurable input; or \item Varying operating conditions that affect the measured parameter but are not related to failure; or \item Trend is not monotonic; or \item Data highly dependent on measurement process; or Data is subject to high levels of process or measurement noise; or \item Reliable confidence limits are required on the extrapolated RUL estimate \end{itemize}\\
ARMA Models \& variants & \begin{itemize} \item Hazard rate is a linear relationship of covariates and noise; and \item Short-term predictions required; and \item Hazard rate is independent of age (i.e., exponential distribution); and \item Measurement data is available for modelling and application but historical failure data is not \end{itemize} & \begin{itemize} \item Hazard rate is not a linear relationship of covariate and noise; or \item When historical or expert data is available in addition to measurement data; or \item Long-term predictions are required; or \item Sufficiently large volume of data is not available for model construction and validation \end{itemize} \\
\end{tabular}
}
\end{table}

\begin{table}[htbp!] %
\centering
{\tiny
\caption*{Table \ref{tab:whentouse} (continued).}
\begin{tabular}{L{2.0cm} L{4.0cm} L{5cm}}
\hline
\hline
Model & Advantages & Disadvantages \\
\hline
\textbf{Statistical} & & \\
PHM & \begin{itemize} \item Times to failure are independent and identically distributed; \item Covariate have a multiplicative effect on the baseline hazard rate; and \item A number of covariates are available and required to describe change in risk; and \item Process represented by covariates is stationary (unless using Dynamic PHM); and \item Associated covariate data is available for the failure modes being modelled; and \item Only the final RUL estimate and confidence limit is required (not an estimate of a precursor to failure) \end{itemize} & \begin{itemize} \item Failures have not occurred previously or have no associated covariate data \item Hazard rate is not multiplicative; or \item Failures cannot be segregated into individual (or dominating) failure modes; or \item Covariates related to the failure modes being modelled cannot be measured; or \item Process represented by the covariates is non-stationary; or \item If a precursor to failure is to be predicted rather than final failure itself \end{itemize}\\
\end{tabular}
}
\end{table}

\begin{table}[htbp!] %
\centering
{\tiny
\caption*{Table \ref{tab:whentouse} (continued)}
\begin{tabular}{L{2.0cm} L{4.0cm} L{5cm}}
\hline
\hline
Model & Advantages & Disadvantages \\
\hline
\textbf{Artificial Neural Networks} & &\\
For Forecasting with ANNs & \begin{itemize} \item Large amount of noisy, numerical, temporal data; and \item Physical, statistical or deterministic model is not known or impractical to apply; and \item An exact optional answer for RUL is required \end{itemize} & \begin{itemize} \item Data is complex or symbolic; or \item Justification or physical extrapolation not required; or \item Temporal inputs are not available; or \item Minimal data is available for training \end{itemize}\\
Parameter Estimation with ANNs & \begin{itemize} \item An RUL model (typically a physical model) is available but contains unkown parameters; and \item Large amount of noisy, numerical temporal data; and \item An exact optimal answer for RUL is required \end{itemize} & \begin{itemize} \item Data is complex or symbolic; or \item Minimal data is available for training \end{itemize}\\
\hline
\textbf{Physical models} &&\\
Physical Models & \begin{itemize} \item Failure modes are well understood and defined; and \item A physical model for each failure mode is available; and \item Operating conditions can be monitored and statistically represented; and \item Process/condition data is available; and \item High-accuracy and precision required in RUL prediction \end{itemize} & \begin{itemize} \item A physical model is not available \end{itemize} \\
\end{tabular}
}
\end{table}

\newpage
\section{Life expectancy models}
The basic idea in developing the life expectancy models is to determine the RUL of a component with respect to the expected risk of deterioration. It is also assumed that the operating conditions are known.
\subsection{Stochastic models}
Stochastic models provide reliability-related information, such as Mean Time to Failure (MTTF) as probabilities of failure with respect to time. Stochastic behaviour is at the heart of these methods and they are based on the assumption that the times to failure of identical components can be represented by statistically identical and independent random variables and thus be described by a probability density function. One main driving factor of these models is the existence of data, which in the case of sparse failures leads to overly pessimistic  estimates. It is shown that the accuracy of the estimate of MTTF can be improved by utilizing censored (suspended data) (times at which failure has not occurred or there is no evidence of failure) \cite{jardine2013maintenance}. The ability to use censored data is important since most of the experimental data is attained through accelerated tests using experimental rigs or bench tests and most of the time the accelerated tests are terminated after a certain period of time and consequently results in censoring. Using censored data is not necessarily helpful especially in small data sets in which censoring might occur early in life and this can introduce other errors \cite{carter1986mechanical}. In the simplest form of application, RUL is equated to the time remaining before a critical number of failures (e.g., 5\%) are expected to occur.

\subsubsection{Aggregate reliability functions}
This is the standard approach widely accepted and used in industry, especially in certain problems for which reliable and considerable amount of data is available. Detailed information on applying statistical distributions to modelling and failure data can be found in various publications \cite{carter1986mechanical, crowder1994statistical, rausand2004system, todinov2005reliability, smith2011reliability, blischke2011reliability, o2011practical}. The overall task consists of determining a probability density function and its related hazard function for a population of components and analyzing the time to failure (TTF). Obviously, the density function is the representative of the whole population and not a single fault progression. In the simplest theoretical approximation, a fault progression curve typically follows an exponential curve and provides information about the expected time of failures. There are various mathematical relations to approximate the probability distributions that best model the failure data (e.g., Exponential, Gaussian, Normal, Lognormal and Weibull functions). Gaussian distribution is the most famous and commonly used distribution in reliability engineering due to its ability to describe many different failure types. The classical well-known bathtub curve (see Fig 7 in \cite{sikorska2011prognostic}) is most commonly described as a piece-wise function made up of three Weibull distributions, each of which describes a different set of dominating failure modes, i.e., early (infant failures), random failures and wear-out failures.

For more complex systems there exists another model for reliability estimation assuming that load and material strength distributions are known. This model is known as Overstress Reliability integral \cite{todinov2005reliability}. Failure data can be fitted to a Weibull distribution using a variety of parameter estimation methods, such as least squares, moments and maximum likelihood. These models make the most famous distributions and are usually incorporated in commercially available softwares. All of these models still depend on reliable large sample sets of failure data points, which have to be collected and stored during extensive (time consuming) tests or under real environmental conditions. In addition, any situation where the failure distribution is exponential, reliability analysis on its own is insufficient for estimating RUL. This is due to the fact that the hazard rate of an exponential distribution is constant over the life of a component and is independent of its service life.

On the other hand, what makes the reliability-based modelling approaches appealing is that distributions are usually derived from observed statistical data and are mathematically easy to construct. The required data can often be extracted relatively easily from a company's existing computerized maintenance management systems. In addition, they provide confidence limits for the results, which is an essential information for decision making. Consequently, analysis of the results is also relatively straightforward and can be performed by reliability engineers and avoids expertise on the subject under study. From a theoretical standpoint, the reliability analysis can be extended to include larger systems by combining the failure data appropriately. In practice however, it is not advised to aggregate too many failure modes together since the failure distributions of a system behaves similar to that of an exponential distribution, which is problematic as discussed earlier. More advanced prognostic models are required for estimating RUL of systems.

\subsubsection{Conditional probability models}
A number of stochastic models try to use conditional reliability functions in conjunction with the Bayes' theorem. In essence, a conditional reliability function is used to describe the current state of the component. The future behaviour/status of the component is estimated based on the recursive update of the conditional function through direct or indirect utilization of Bayes' theorem (thereby they could also be referred to as Bayesian models). Knowing the current state of the asset, once a conditional reliability function is determined, the RUL function is defined as the conditional expected time to failure (which may or may not be time dependent) \cite{oakes1990note, maguluri1994estimation, reinertsen1996residual, yuen2003mean}. Modelling variants differ in the calculation procedure of the conditional probability function as well as the kind of information used to define the current state.

\subsubsection{RUL probability density function}
The RUL probability density function is probably the simplest Bayesian approach which is an extension of traditional aggregate reliability analysis.  It requires the probability density function of the relevant failure mode. Information is then obtained to locate a specific item on this general distribution (e.g., an age at which the item has not failed). This population grows in size as the new data is appended and similarly the distribution is amended to consider this information using Bayes' theorem \cite{lewis1986optimal}. The process repeated each time a new data point is available and this process is called Bayesian `updating'. There are various names to the resulting distribution, i.e., the predictive density function or the remaining RUL PDF. It is also possible to derive a credibility interval (equivalent to a confidence interval) \cite{rausand2004system,clarotti1989bayes}. It is also possible to make improved predictions for the new state (i.e., the condition probability, or posterior function) by incorporating more advanced state estimation techniques such as Kalman filtering, particle filtering methods. The rational behind selection of the most appropriate method depends on both the system as well as the noise type. In addition, predictions for the next state often involve evaluation of integrals that do not possess closed-form solutions. Thus integration approximation methods are often required, such as regression models \cite{maguluri1994estimation} or bootstrapping methods \cite{yuen2003mean}, to estimate the expected value and covariance of these PDFs. The accuracy and precision of RUL estimation using this technique improve as the end of life approaches. Besides, it is also relatively simple to calculate and use these techniques.

\subsubsection{Static Bayesian Networks}
Bayesian Networks (BN)/Bayesian Belief Networks (BBF) are probabilistic acyclic graphical models that represent a set of random variables and their probabilistic interdependencies. Depending on the type of the information used, these can also be considered as either knowledge-based, stochastic or hybrid approaches. There are a number of nodes, which are connected by directional arcs that represent a direct causal influence between nodes in a mandatory acyclic pattern. The nodes themselves can take on distinct states or levels and represent random variables. The strength of the causal influences are quantified using conditional probabilities. Ultimately, each node has a conditional probability table that defines probabilities for each state of the node given the states of its parents \cite{dey2005bayesian}. Given a network design configuration and nodal conditional probabilities, a BN can be used to evaluate the likelihood of each possible cause being the actual cause of an event. It could also represent probabilities associated with a particular event occurring next if time series modelling is adopted. The output of the BN is in the form of probabilities, which intrinsically contain information about their confidence. This is a major advantage. A detailed mathematical description of BN modelling in reliability and a list of modelling software available can be found in \cite{langseth2007bayesian}.

\subsubsection{Dynamics Bayesian Networks}
Dynamics Bayesian networks are those in which the directed BN arc flow forward in time and are therefore useful for modelling time series data \cite{ghahramani1998learning}. Prognostic URL estimation is invariably undertaken using time series forecasting as in \cite{weidl2003object}. The most common variants used in engineering prognostics include Markov models, Kalman filters and Particle filters. For a detailed review of Markov models see Ref. \cite{sikorska2011prognostic}.

\subsubsection{Bayesinan estimation with Kalman filters}
Both Kalman and Particle filters (which is discussed in the next section) are not different types of models, but rather different approaches to implementing generic dynamic BNs. Howevver, they are widely used in engineering prognostics and deserve particular attention, which requires a brief overview of the underlying assumptions, limitations and strengths of these specific approaches. The complexity of the dynamics and the type of noise are crucial in assessing the domain of the application of these methods. The Kalman filter is a computationally efficient recursive digital processing technique used to estimate the state of a dynamic system from a series of incomplete and noisy measurement in way that minimizes mean squared error. It is the most famous estimation method within the control community. At any instant, it is defined by its state estimate and error covariance. In five steps, it estimates unknown states from only current observations and the most recent state and these states need not be directly measurable \cite{welch2006introduction}. Kalman filtering assumes certain features for the process and measurement noise i.e, Gaussian, white, independent of each other and additive. Traditionally, it was also assumed that the dynamic being modelled needed to be linear, however, it has been shown that this is not the case if the aforementioned assumptions on noise holds \cite{ito2000gaussian, haug2005tutorial}. During the iterative procedure, it is necessary to solve a number of integrals; but if the linearity assumptions are met, they have exact solutions and it is not necessary to use approximation methods.

The Kalman filter requires an appropriate initial quantification of the measurement noise covariance, which is relatively easy when observations are stationary. However, determining the process noise covariance is more challenging as it is often not possible to directly observe the process being modelled. The performance of the filter improves when these parameters are tuned separately to their proper values. The filter will reach steady-state very quickly if both noise covariances are constant between iterations, i.e., process and observed data are stationary. There are several variants of Kalman filter. For instance, Extended Kalman filter (EKF) is a modification of basic Kalman filter free of the assumption regarding the linearity of either the underlying process or of the relationship between the process and the measurements. Instead, partial derivatives of the process and measurements functions are calculated to linearize the estimation around the current state prediction. Unfortunately, this also transforms the noise, which no linger remains Gaussian, thereby invalidating one of the filter's original assumptions. This is a fundamental flaw in the EKF model, the effect of which is that the state estimator only approximates the optimality of Bayes' rule by linearization \cite{welch2006introduction}. It also requires a solution (albeit approximate) for a Jacobian matrix, which is difficult to find. Computationally, it is less efficient and process time increases as all covariance and model parameters need to be recalculated in each iteration. Most importantly, there exists the possibility of filter divergence. Traditionally, the EKF was the most popular Kalman variant for state estimation of non-linear processes. However, due to the issues mentioned, and improvement is computational resources alternatives have recently been developed \cite{ito2000gaussian, haug2005tutorial}. The Gauss-Hermite quadrature Kalman filter (GHKF), a modified version of the GHKF called the unscented Kalman filter (UKF), and Monte-Carlo Kalamn filters (MCKF) are all variants of the basic Kalman filter applied to non-linear processes; they differ in how estimates for the Kalman filter integrals are calculated and consequently have varying computational efficiencies. For all of the variants, assumptions about Gaussian noise are still required \cite{haug2005tutorial}. In practice, if the system has a large number of states, the UKF is the technique of interest for especially when the non-linear functions are smooth \cite{haug2005tutorial, guo2006quasi}. Specific examples of applying Kalman filters to BN for the purposes of RUL estimation of engineering assets can be found in \cite{swanson2001general, ray1996stochastic}.

\subsubsection{Bayesian estimation with particle filters}
Particle filters are the candidate alternatives to Kalman filters as they are not constrained by linearity or Gaussian noise assumptions. They are particularly useful for situations in which the posterior distribution is multivariate or non-standard. The principle different between Kalman filter and Particle filter (with respect to how they calculate the posterior PDF) is that the former relies on extrapolating from the prior state, whereas the latter uses a sequential importance sampling scheme to simulate the entire next state in every iteration of the filter. Particle filter does this by generating a set of random samples (also known as particles) from a theoretical density function and then adjusts the associated set of particle weights at each iteration. Samples of dynamic noise are also generated with each cycle. It is important to note that with sufficient samples, Particle filters are more accurate than either the EKF or UKF. In addition, compared to the classical Monte-Carlo integration, they require fewer samples to adequately approximate the distribution, which results in a superior computational performance. However, there are certain problems in real applications. The first difficulty is that as the number of iterations increases, the filter can degenerate and the posterior PDF approximation becomes zero \cite{haug2005tutorial}. One possible method of avoiding this problem is obviously to increase the number of samples and reduce the number of iterations. Unfortunately, this is not always practical due to the increased computation time. Alternatively, a re-sampling step can be introduced to each time interval that replaces low probability particles with the same number of high probability particles. A number of different re-sampling methods can be used \cite{evensen1994sequential, carlin1992monte}, including the inverse transformation method \cite{haug2005tutorial} and the Bootstrap Particle Filter \cite{franco2002comparison, rodriguez2009bootstrap, gordon1993novel, stoffer1991bootstrapping}. A detailed discussion on optimal sampling (and re-sampling) is given in \cite{djuric2002perfect}. There are also a number of Particle filter approximation techniques that do not involve re-sampling. These use Monte-Carlo, Gauss–Hermite or Unscented Kalman filters to define an importance density functions, from which particles are sampled. In the first two of these methods the importance density is assumed to be Gaussian, based on the mean and covariance output of the updated prior density. According to Haug, both the Gauss–Hermite and Unscented Particle filters work well and are implementable in real-time, while the Monte-Carlo Particle filter requires an excessively large number of samples and outliers can result in numerical instabilities that prevent convergence \cite{haug2005tutorial}. Although particle filters have been used extensively in both econometrics and target trajectory forecasting, there are only a few published example applications related to asset health prognosis \cite{orchard2005particle, cadini2009model}. \cite{orchard2005particle} used sequential importance sampling Particle filters to estimate the time progression of a fatigue crack, which was modelled with a combined state dynamic model and a measurement model to predict the posterior probability density function of the stage (the fatigue crack growth) \cite{orchard2005particle}. Similarly, in \cite{cadini2009model} a combined Bayesian-behavioural model is used along with Particle filters for prediction of fatigue crack growth progression.

\subsection{Statistical models}
In this section we talk about the statistical branch of Fig.~\ref{fig:URLModels}. Statistical models use previous inspection results on similar components to estimate both initiation and progression of a possible failure mode. They are most of the time used in problems when suitable dynamic model is not available as an alternative for ANN. The nominal (standard) behaviour of the component is used as a reference and prediction of the RUL is achieved by comparing the current behaviour to the nominal one. Statistical models are generally categorized among the data-driven methods since they utilize temporal data such as condition or process monitoring outputs.
 
\subsubsection{Trend evaluation}
Probably the simplest approach to RUL prediction is based on trend analysis. It uses the trend of a single monotonic parameter which is believed to be related to the remaining life of the component. The selection of the appropriate ``feature parameter'', which may represent a single sensor (feature) or a number of sensor (features) is critical. This one feature parameter is then plotted as a function of time and is used along with a pre-defined alarm level. A warning end-of-life signal will be triggered when the feature parameter reaches the alarm level. In fact, there could be several alarm levels depending on the severity of the health deterioration of the component and its capability to perform the assigned tasks. For instance, there could be alarms for early-warning and `final failure' (denoted by asterisk and red circles in Fig.~\ref{fig:FailureModes}). Standard regression methods are used to calculate a candidate trend e.g, polynomial fit using a least-square method. There are situations in which there is no data for all parameter levels up to and exceeding the alarm limits and requires trend extrapolation. However, in engineering prognosis, often times failure mechanisms change with the progress of failure and may alter the trends significantly. Therefore, interpolation is always preferred over extrapolation. It is also important to chose alarm limits with acceptable accuracy (through data records and personnel knowledge). It is obvious that if a somewhat conservative limit is selected, there is great possibility of premature replacement, and in contrast, if a high value for alarm is used, it is highly probable that the algorithm will miss a failure. With respect to the confidence limits, if only interpolated data is being used, confidence levels on the prognosis can be calculated based on the variance of the underlying trend. However, confidence limits cannot be calculated for extrapolated regions. Implementation of this type of trend evaluation is simple and easy but indications of impending failure are typically noisy and often non-monotonic \cite{engel2000prognostics}. The failure situation gets even more complicated when multiple failure modes exist. Consequently, simple thresholds may not result in a reliable RUL prediction, particularly where data needs to be extrapolated. On the other hand, as the failure is approached damage conditions become clearer, which results in clearer trends. Note also that parameters most appropriate for predicting RUL may not be the same as those used for detecting the beginning of a fault mode. For example, in the early stages of bearing's failure, Kurtosis (4th statistical moment) of a bearing's vibration signal often increases but can then decrease as the bearing approaches its end of life.

\subsubsection{Autoregressive models}
Forecasting of time series data are widely achieved through using Autoregressive moving average (ARMA), Autoregressive integrated moving average (ARIMA) and ARMAX models \cite{box2015time}. In all the variants, a linear function of past observations (and random errors) is used to calculate the future value; a comprehensive summary is presented in \cite{lewis1992applied}. The three mentioned autoregressive models are slightly different in the linear equation, which is used to relate inputs, outputs, and noise. ARMA and ARMAX models should only be used for stationary data since they can remove temporal trends. Note that a time series is defined to be (weakly) stationary when its first two moments, i.e., mean and variance, respectively, are time-invariant under translation \cite{tsay2000time}. The autocorrelation also needs to be independent of time. Consequently, prior to modelling it is essential to perform trend tests to ensure the validity of the stationarity assumption. ARIMA models, that use the concept of integration enforcement, are capable of describing systems with low frequency disturbances. Autoregressive models are developed in three recursive steps:
\begin{enumerate}[I.]
  \item Model identification: Initially, using a set of time series data, values for the orders of the autoregressive and moving average parts of the ARMA/ARIMA equations are hypothesized, as well as the regular-difference parts for the ARIMA model. A suitable criterion of fit is also assumed.
  \item Parameter estimation: Using non-linear optimization techniques (e.g., a least-squares method), parameters of the ARMA/ARIMA equations are calculated to minimize the overall error between the model output and observed input-output data.
  \item Model validation: A number of standard diagnostic checks are used to verify the adequacy of models, utilizing unseen data. According to \cite{wu2007prognostics}, options include the following: examining standardized residuals, autocorrelation of residuals, final prediction error, Akaike information criterion, and Bayesian Information Criterion (BIC).
\end{enumerate}
These three steps are repeated until a satisfactory model is obtained. Once the model parameters are fixed, it can be used to forecast future values; if the minimum mean squared error is used as the criterion these are simply the conditional expectations of the model. However, note that, typical ARMA models (and variants) are effective for short-term predictions, but less reliable when used for long-term predictions. They are not reliable for the long-term predictions due to dynamic noise, their sensitivity to initial system conditions and an accumulation of systematic errors in the predictor \cite{box2015time}. An extension of the basic ARIMA approach is proposed in \cite{wu2007prognostics} that uses bootstrap forecasting for machine life prognostics. This variant avoids  using previous values predicted to forecast future values, and instead generates predictions only based on true observations. As parameters were updated in realtime, the model was able to adapt to dynamic changes in the operating process and did not suffer from error accumulation. Predictions were superior to those based on traditional ARIMA models. Another example utilizing ARMA modelling for prognostic estimation is presented in \cite{yan2004prognostic}. Although few details are available on the models themselves, ARMA techniques have also incorporated into the prognostic and data fusion software developed by the NSF Center for Intelligent Maintenance System as described in \cite{lee2006intelligent} (the system also uses other types of modelling for residual life estimation including proportional hazards and neural network approaches). One novel alternative to ARMA methods for prediction of time-series data worthy of mention uses Dempster–Shafer regression. Application of this technique to machinery prognostics is presented in \cite{niu2009dempster}. It offers significant potential for applications where temporal trends of prognostic parameters (to be used for extrapolation) are non-linear and/or chaotic and thus can not be modelled using ARMA techniques. Reference~\cite{tang2015optimal} discusses the application of autoregressive to slowly degrading systems subject to soft failure and condition monitoring at equidistant, discrete time epochs.

\subsubsection{Proportional hazards modelling}
Proportional Hazards Modelling (PHM) was first proposed in \cite{cox1992regression} and models the way explanatory or concomitant variables, also referred to as covariates, affect the life of the equipment, and at the same time, is one of the most extensively used models for prognostics. The basic difference between PHM and the linear regression methods is that the former assumes a multiplicative relationship for covariates, whereas that latter assumes an additive effect on the overall hazard rate. PHM models deterioration as the product of a baseline hazard rate, and a positive function. The multiplicative function reflects the effect of the operating environment on the baseline hazard and is described by a vector of covariates and an associated vector of unknown regression parameters. RUL can be deduced from the associated survival function \cite{cox1984analysis}. The positive function is usually assumed to be exponential (primarily for convenience) although other mathematical functions such as logarithmic, inverse linear, linear or quadratic functions are also common \cite{kumar1994proportional, elsayed2007design}. It is possible for the elements of the covariate process vector to take positive values implying that the covariate is actually improving the condition and thus reducing the hazard rate when compared to the baseline. For instance, increased corrosion inhibits (or concentration reduces) the rate of internal corrosion \cite{kumar1994proportional}. For some mechanical as well as electrical components the seasonal variation of temperature results in positive and negative contributions. Covariates are often referred to as internal or external. In the prognostics context, internal covariates refer to outputs generated by the component being degraded and thus only exist as long as the degraded component remains in service. An example of internal covariate is the vibration level at the inner race bearing frequency that is used to predict bearing failure. Internal covariates can also be considered `response covariates' as they are generated in direct response to the failure process. On the other hand, external covariates refer to outputs generated by an independent process; they can also be considered `risk factors' and are usually not affected by repairing or replacing the degraded component. An example for external covariate is sulphur concentrations in crude oil that may be used to indicate increased risk of process pipe corrosion. A more detailed discussion on covariates is given in \cite{kalbfleisch2011statistical}. PHM works according to a number of assumptions:
\begin{enumerate}[a)]
\item Times to failure are independent and identically distributed, i.e., perfect repair.
\item Covariates have a multiplicative effect on the baseline rate.
\item Individual covariates are independent (i.e., the value of the covariate function for one item does not influence the time to failure of other items).
\item The effect of the covariates is assumed to be time independent.
\item All influential covariates should be included in the model.
\item The ratio of any two hazard rates is constant with respect to time (thus the respective survival curves will not intersect).
\end{enumerate}

It is possible to use graphical analytical goodness-of-fit tests to verify whether these assumptions are valid for the system being modelled \cite{kumar1994proportional, ansell1997practical}. Applying PHM requires the estimation of the parameters of the baseline hazard function and the covariate process vector. Initially, the covariate vector weightings are calculated irrespective of the form of the baseline hazard function for which the maximum likelihood method is the most commonly applied technique, but a variety of other approaches have also been used \cite{bendell1985proportional,cox1992regression,kumar1994proportional,mazzuchi1989assessment,makis2003optimal}. Higher weightings are given to the covariates that are good indicators of failure, while those with little correlation to failure are assigned much smaller weightings. The accuracy of the modelling is improved if only relevant covariates are incorporated into the model. Consequently, a backward step wise procedure is often implemented to exclude the least significant covariates and re-estimates the model parameters; this procedure is repeated until all remaining factors are significant. Alternatively, new variables can be sequentially forced into the model during the search for significant covariates. Once covariate parameters have been defined, variables of the baseline hazard function can be estimated using either parametric or non-parametric methods. The latter is generally preferred by statisticians as the form can be estimated from the data \cite{bendell1985proportional,cox1992regression,kumar1994proportional,dale1985application}, which is then compared with various standard distributions forms to identify the most appropriate model. In practice, however, the baseline hazard function is often assumed in advance to be a Weibull or exponential function to facilitate the use of common parametric regression methods. Due to the confusing effects of the covariates, these pre-assumed forms of the hazard function may not be justified and may not be the best choice \cite{bendell1985proportional}. To ensure that the baseline hazard function remains physically meaningful it is desired to configure covariates in a manner that they equate zero for the `baseline' operating state (although this requirement is not a mathematical constraint). A critique of early attempts to apply proportional hazards model to problems of engineering reliability was provided by \cite{bendell1985proportional}. A more comprehensive review is conducted in \cite{kumar1994proportional}. Nevertheless, the body of work on PHM clearly demonstrated the advantages of PHM over standard regression techniques, including the ability to manage
nuisance variables (unrelated covariates), censored data \cite{dale1985application}. Over the past years, these basic techniques have been refined, extrapolated and expanded, particularly for the purposes that include optimizing maintenance decisions \cite{makis1995optimal,jardine1989proportional,makis1991optimal, liu1997joint, jardine1999optimizing, jardine2001optimizing, jardine2002optimizing, vlok2002optimal, kumar1994proportional, samrout2009optimization}, analysing data obtained from accelerated life tests \cite{elsayed2007design}, and appliction to model systems that are subject to partial repair \cite{makivs1991optimal,liu1995replacement,jiang2001optimal}. Specific examples of industry led research include \cite{dale1991assessment,krivtsov2007recent, drury1988proportional, baxter1988proportional}. 

A variation of PHM that does not assume perfect repair, known as the Proportional Intensity Model, exists and has also been applied by several researchers \cite{lugtigheid2007optimizing}. Intuitively, the best models are expected to be based on a mixture of diagnostic indicators, which seems to be problematic due to the lack of published examples. To overcome the problem of insufficient failure data, Ref \cite{mazzuchi2008paired} has applied an expert judgement approach (paired comparison), in conjunction with a small amount of actual failure data to populate the PHM parameters \cite{mazzuchi2008paired}. Collectively, this work to date on the application of PHM to asset prognostics has been conducted using highly selective and well-controlled data sets (albeit some of the data was collected from real operating plants); in each case, only a small number of overlapping failure modes was modelled. Thus, the ability of PHM to estimate RUL for prognosis of varied faults in complex systems is uncertain and likely to be an ongoing challenge. The extension of the PHM to complex repairable systems with a number of sub-systems is a difficult task. A complex system has several components with their associated failure modes and the assumption that failures are independent and identically distributed is far from truth. In addition, it is hard to find a comprehensive set of covariates to describe all failure modes. However, as equipment becomes more reliable it is difficult, from a practical standpoint, to obtain sufficient data pertaining to failures and corresponding covariates to model all failures \cite{raouf2006data}. Furthermore, data aggregation can obscure information about component failures thus making it difficult to produce consistent and applicable histories \cite{ansell1997practical}. All in all, it is suggested to apply PHM at the failure mode level when appropriately refined failure histories and physically relevant covariates are known. This is not a straightforward task as it requires more data collection procedures in addition to the knowledge of  the physical root causes of failures (i.e., the failure mode). In addition, associated working age and diagnostic information must be recorded accurately \cite{vlok2002optimal} and in an accessible format. 

Note that given current modelling limitations, when a subsequent repair/replacement is made, suspension events need to be recorded against all other potential failure modes that are affected by that repair/replacement, so that working ages for these failure modes can be adjusted. (This
is only practicable if implemented as an autonomous process.) This is done to avoid biases in estimates for the RUL, which may result in underestimations \cite{vlok2002optimal}. More information about data requirements for PHM is given in \cite{raouf2006data}. Since the PHM relies on the data of particular failure modes, it is not capable of estimating RUL of failure modes, which have not occurred previously. 

The traditional PHM approach to non-stationary process (e.g., reliability growth of repairable system) is performed by using a stratified approach \cite{dale1985application,prentice1981regression} in which data is grouped based on various failure times in the life cycle for each component. Each step is then modelled individually, with required failure histories for each step \cite{kumar1994proportional}. Dynamic PHM is an alternative to the traditional PHM where a generalization of the basic PHM equation is used to take into account time dependent covariates \cite{makis1991optimal,makis1991computation, makis2003optimal}. The dynamic model is capable of predicting the future development of covariate and failure times. In this work, the non-stationarity was accommodated by assuming that the covariate vector was a multivariate non-homogenous Markov process with all but failure state hidden. It was also assumed that covariates were only observable at certain times (i.e., periodic inspection/monitoring) \cite{makis2003optimal,makis1991optimal}. Backward recursion algorithms and optimal stopping framework were then used to determine optimal inspection intervals (based on minimizing expected cost) as well as to calculate the expected RUL. It was assumed that the system was renewed after replacement. In practice, only a few states are needed to represent the failure process, such as those corresponding to periods of `wear-in', `normal' and `wear-out'. Some publications have reported the results of this approach for analyzing real operating systems \cite{jardine1999optimizing,vlok2002optimal,banjevic2001control}.

\subsection{Physical models}
Physical models (also known as physics of failure or behavioral models) exploit physical laws to quantitatively characterize the behaviour of a failure mode. Obviously, this requires a thorough/detailed understanding of the system behavior in response to external loads such as stress, at both macroscopic and microscopic levels. It is based on the fact that it is possible to describe the behavior of a component accurately and analytically. At the heart of the method is the estimation of an output for the RUL of a component by solving a deterministic equation or set of equations derived from extensive empirical data. Data includes common scientific and engineering knowledge as well as those acquired through specific laboratory or filed experimentation. The physical model entails physical properties, constant parameters of the equations, corrosion rates, etc. In the end, the model is described by using a series of ordinary or partial differential equations that can then be solved in most cases with Lagrangian or Hamiltonian dynamics, approximation methods applied to partial differential equations, distributed models \cite{vachtsevanos13intelligent, roemer2000advanced}. A more comprehensive list of behavioral models is provided in \cite{heng2009rotating}.

Once a model is developed and verified, sensor measurements of the actual component are used against outputs of the developed model to calculate the residuals (i.e., differences between reality and the model); The status of fault is drawn if large residuals are observed while small residuals are attributed to noise and modelling errors under normal operating conditions \cite{luo2003model}. A number of thresholds could be defined to identify the presence and/or condition of faults. There are several methods to calculate the residuals including parameter estimation, state-space methods or parity equations the benefits of each are discussed in \cite{isermann2006fault}. The projection of the degradation behavior into the future is used for estimating RUL. One has to define characteristics for a set of features and their associated levels of accuracy in order to construct a physics of failure model:
\begin{enumerate}[a)]
\item Identify likely initiating failure modes for which behavioral models are required.
\item Process behavior across possible/typical operating ranges.
\item Degradation behavior under aforementioned process conditions.
\item Relationship between process measurements and degradation behavior(s).
\item Process and measurement noise.
\end{enumerate}

In practice and for many cases, the above parameters are inherently probabilistic random variables thus demands the incorporation of their statistical distributions into the model. This enables the estimation of confidence limits. For the purposes of prognostics, failure mechanisms can be broadly divided into two categories \cite{dasgupta1991material}. The first type of failure is associated to over stress failures and they occur when incurred load exceeds the strength of the material; this kind of stress is not destructive, i.e., they have no long-term effect once the load has been removed. Examples include brittle fracture, yielding and buckling. Once the particular damage reaches the allowed tolerance, normal operating loads exceed the remaining strength of the material and an over-stress failure is said to have happened \cite{blischke2011reliability}. The second type of failure is wear-out failures which are characterized by accumulated damage that does not disappear when the load is removed (e.g., fatigue, wear in brake pads). A physical prognostic model for wear-out failure modes needs to be able to track aggregated damage and its rate of progression under any/all operation conditions. 

Consequently, if available and when sufficiently complete, physics-based models tend to significantly outperform other types of models \cite{luo2003model}. Additionally, the outputs of physical models are easy to interpret. Their obvious disadvantage is that the behavior of the system must be derivable from first principles, which may not always be possible due to an imperfect understanding of how the failure mechanisms behave under the range of relevant operating conditions. Even if the mechanisms are fully understood, assigning the appropriate parameters for all aspects of the model requires a significant volume of accurate and reliable multivariate data that is rarely available. Condition indicators specific to the failure mode being modelled must also be identified and continually collected. Consequently, physics of failure models tend to be used in isolated cases, for well-understood faults in simple systems and/or by users with established diagnostic systems and predictive maintenance programs. It seems that crack propagation failure modes are the most commonly developed behavioral models for prognostics.

\section{P\&HM Tool Selection Method}
In this section, a procedure is explained to facilitate the selection of the most appropriate methods/algorithms for various steps involved in P\&HM \cite{lee2014prognostics}. The available data is a steppingstone toward developing a systematic approach to apply P\&HM methodologies to traditional as well as novel areas. Prior to examining the possibility of using certain algorithms, it is pivotal and beneficial to understand the characteristics of the data and possible causality between these characteristics, and the nature of the system in terms of the operating condition, service intensity, system dynamics and all other applicable attributes. Achieving an effective methodology with the appropriate blend of algorithms that results in reliable accuracy is, more or less, dominated by the characteristics of the data whether it is in the form of vibration, acoustic emissions, environmental parameters, etc. The major tasks in the field of PHM are already described and include signal processing, feature extraction and reduction, fault diagnosis, health assessment, performance prediction and so on. However, within each task there exist a multitude of algorithms which have been developed and benchmarked to process signals, classify them and using the results determine the current health status and estimate the future condition of the component under study. Practitioners and researchers usually have different algorithms preferences which is dependent upon the application and available infrastructure. Table~\ref{tab:ModelsAdvantages} presents a list of the most commonly used algorithms including their applications, strengths and weaknesses in the field of P\&HM. Algorithm selection is a crucial step for developing an effective P\&HM system, which will ultimately affect the results and their confidence limit.

One way of selecting algorithms is to take a heuristic approach that relies on researchers' experience and expertise to meet users' requirement. However, such an approach is not ideal for situations in which there is a lack of expert knowledge (and expert personnel), and could be time-consuming for complex problems or systems. The goal is to achieve the following targets in the most efficient way: 1) to provide quantified selection criteria, 2) to enable automatic benchmarking, and 3) to recommend the appropriate tool(s) for a particular application. It is therefore inevitable to devise a selection scheme that compares and ranks the suitability of each algorithm by considering the application attributes, proficiency and requirements of the end user. A feasible solution can be a numerical comparison based on the ranking of algorithm scores for each category so that the top algorithms can be selected. A suitable ranking method for algorithm selection is Quality Function Deployment (QFD) \cite{revelle1997qfd} and is best known as a tool for product design, quality management, customer need analysis and decision making purposes \cite{chan2002quality}. In traditional QFD, a House of Quality (HOQ) is constructed to combine engineering attributes and customer needs (and their assigned weights), and transform them into design specifications and controllable parameters. This provides the quantification step of the aforementioned targets. For algorithm selection, data characteristics, corresponding algorithm suitability and user inputs are integrated to give a ranking of all algorithm candidates in each category. The process of applying QFD for algorithm selection can be summarized through the following main steps:
\begin{enumerate}[1)] 
\item According to the application and available knowledge of the data, all criteria related to the application should be selected. 
\item The properties of each criterion should be identified in order to describe the available data in a detailed manner. These criteria do not have to be binary as long as ambiguity can be avoided, so users can classify the level of the characteristic with high confidence. Furthermore, a quantified description is recommended in this step, for example, low, medium and high stationarity can be presented by ascending integers like 1, 3 and 5. 
\item Eligible algorithms for each characteristic are compared in a pair-wise way based on algorithm applicability. Analytical Hierarchy Process (AHP) \cite{bhushan2007strategic}, employed in this step, is a decision making tool, which is able to compare every paired combination of algorithms and provide final applicability indices for all algorithms for each specific characteristic.
\item An HOQ can be established for each one of the algorithm categories to aggregate all the indices from the previous step for all characteristics in order to generate an overall weight, based on which a final rank of algorithms can be decided for this category. Therefore, a procedure can be established for the user to execute an effective, systematic P\&HM approach with the top ranked algorithms from each category.
\end{enumerate}
As an example application, Figure 5 in Ref \cite{lee2014prognostics} illustrates the QFD algorithm selection tool using the gears of a wind turbine. Rotary components of a wind turbine system such as rotor blades, bearings, shafts and gears are working under dynamic loads and are more susceptible to failure than other components. The aforementioned systematic methodology can be referenced to explore and develop fundamental techniques to aid in establishing a P\&HM system for wind turbines under varying environmental, operational and aging processes. In this case, only vibration data is available, and the applicability of each relevant algorithm is defined by assigning an importance to different characteristics using a scale ranging from 1 to 5 (5 being the most important). The rankings are structured so that in each catalog the algorithm with lowest ranking is the most recommended one. The tool selection for each of the other critical components can be performed in this fashion.

\subsection{Visualization tools}
After the selected algorithms have digested the data, prognostics information is ready for further utilization to support the decision making process. One valuable objective of PHM is to enable a support system to convey the right information to right person so that judicious decisions can be made at the right time. Therefore, visualization tools are essential parts of a PHM methodology. Four frequently used visualization tools, Degradation Chart, Performance Radar Chart, Problem Map and Risk Radar Chart, can be designed to present prognostics information as shown in below Figure taken from \cite{lee2014prognostics} for the sake of discussion.

\begin{figure}[htbp!]
\centering
\includegraphics[width=6.0in]{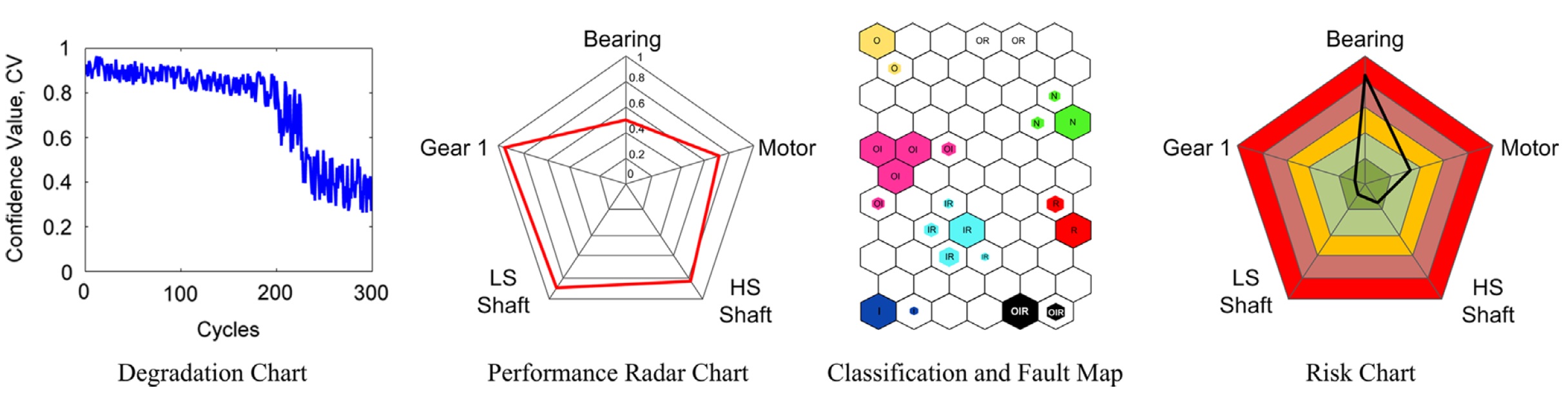}
\caption{Four visualization tools for P\&HM \cite{lee2014prognostics}.}
\label{fig:PHMtoolselection}
\end{figure}

The functionalities of the presented visualization tools are described as follows \cite{wang2009collaborative}:
\begin{itemize}
\item \textbf{Degradation Chart}-If the confidence value (0-unacceptable, 1-normal, between 0 and 1-degradation) of a component drops to a low level, a maintenance practitioner can track the historical confidence value curve to find the degradation trend. The confidence value curve shows the historical, current and predicted confidence value of the equipment. An alarm will be triggered when the confidence value drops under a preset threshold.
\item \textbf{Performance Radar Chart}-A maintenance practitioner can look at this chart to get an overview of the performance status of each component. Each axis on the chart corresponds to the confidence value of a specific component.
\item \textbf{Classification and Fault Map}-A Classification and Fault Map is used to determine the root causes of degradation or failure. This map classifies different failure modes of the monitored components by presenting different failure modes in clusters, each indicated by a different color.
\item \textbf{Risk Chart}-A Risk Chart is a visualization tool for plant-level maintenance information management that displays risk values, indicating equipment maintenance priorities. The risk value of a machine (determined by the product of the degradation rate and the value of the corresponding cost function) indicates how important the machine is to the maintenance process. The higher the risk value, the higher the priority given to that piece of equipment for requiring maintenance.
\end{itemize}

\section{Performance Metrics for Evaluating Prognostic Predictions}
Performance criteria on metrics determine the adequacy of a prognostic approach for a given application \cite{saxena2010metrics}. Extensive work has been reported to define the appropriate performance metrics for a given application and health management and conation monitoring approaches \cite{pecht2008prognostics,saxena2010metrics,coble2008prognostic,wheeler2009survey}. As the literature shows, there are three major performance indicators to determine the efficiency and effectiveness of PHM applications: prognostic distance, accuracy, and precision. The time between the predicted time of incipient failure and actual component failure is called the prognostic distance. This definition of prognostic distance has been derived from the application of canaries \cite{wang2012economic}. Accuracy means the correctness of the remaining life estimates. The correctness of the prediction of time determines the accuracy of prediction. Precision accounts for the uncertainty estimates in remaining life prediction. The width of the uncertainty band determines the precision of the estimates. A shorter band has higher precision, and a wider band has lower precision. The parameters that are of interest to risk-informed applications include assessment of the reliability/safety margin for case or scenario being evolved. Even though extensive work has been performed on the development of performance metrics, there is need for further research on the development of acceptance criteria for performance metrics \cite{sharp2013simple}.

%% file: Application-Challenges.tex
\subsection{Applications}
In this section, we present a list of works on the rotary machinery prognostics as well as other critical components that are widely used in engineering platforms, engines and mechanical systems. Table 5 gives an introductory summary of tools for common critical components, regarding the components' issue and possible failure modes, characteristics, common available data types, common features and algorithms applied for diagnostics and prognostics. Note that the following abbreviations are used for Fourier Transform (FT), Short Time Frequency Transform (STFT), Wavelet Transform (WT), Empirical Mode Decomposition (EMD), Auto-regression (AR), Hilbert-Huang Transform (HHT), Neural Network (NN), Hidden Markov Modeling (HMM), Support Vector Machine (SVM), Genetic Algorithm (GA), Auto-regressive Moving Average (ARMA), Principal Component Analysis (PCA),Wigner-Ville Transforms (WVT), Support Vector Regression (SVR).
\begin{table}[htbp!] %
\centering
{\tiny
\caption{Introductory summarization of tools of critical components.} 

\begin{tabular}{L{1.2cm} L{2cm} L{2cm} L{1cm} L{2cm} L{2cm}} 
\hline
\hline
Item & Issue \& failure & Characteristic & Common measures & Common features & Used models \\
\hline
Bearing & Outer-race, inner-race, roller, and cage failures & Raw data does not contain insightful information; low amplitude; high noise & Vibration, oil debris, acoustic emission & Vibration characteristic frequency, time domain statistical characteristics, metallic debris shape, size, quantity, sharp pulses and rate of development of stress-waves propagatoin  & FT \cite{mcfadden2000application,mechefske1992fault}, STFT \cite{randall2005applications}, WT \cite{qiu2006wavelet}, EMD \cite{lei2007fault}, Bispectrum \cite{yang2002third}, AR Frequency Spectra \cite{wang2008fault}, NN \cite{wang2001fault, yam2001intelligent, huang2007residual}, HMM \cite{ocak2007online,zhang2010hidden}, Fuzzy logic \cite{satish2005fuzzy}, GA \cite{feng2009ga}, ARMA \cite{galati2006application}, Stochastic Model \cite{li2000stochastic,wang2002model}, PCA \cite{zhang2005integrated}\\
\hline
Gear & Manufacturing error, tooth missing, tooth pitting/spall, gear crack, gear fatigue/wear & High noise; high dynamics; signal modulated with other factors; gear specs need to be known & Vibration, oil debris, acoustic emission & Time domain statistical features, vibration signature frequencies, oil debris quantity and chemical analysis & FT \cite{choy1996analysis}, STFT \cite{kar2006technical,bartelmus2009vibration}, WT \cite{peng2004application,suh1999machinery}, EMD \cite{loutridis2004damage,wang2007gearbox, liu2006gearbox}, HHT \cite{liu2006gearbox,li2006wear,brie1997gear}, NN \cite{dellomo1999helicopter, staszewski1997classification,byington2004data,khawaja2005reasoning}, Fuzzy Logic \cite{dempsey2004integrating}, Neuro-Fuzzy Hybrid Model \cite{wang2004prognosis}, Energy Index Analysis \cite{wang2004prognosis}, Kalman
Filter \cite{wu2004application, wu2004application, houser2002comparison}, SVM \cite{samanta2004gear}, Autoregressive Model \cite{wang2002autoregressive,chen2006detecting}, Particle Filter \cite{orchard2009particle} \\
\hline
Shaft & Unbalance, bend, crack, misalignment, rub & Vibration signal is relatively clean and harmonic frequency components of rotating speed can indicate the defects & Vibration & Vibration characteristic frequency, time domain statistical characteristics, system modal characteristics & FT \cite{san2007virtual}, WT \cite{lingli2005research}, Wigner–Ville Transforms (WVT) \cite{kim2007comparative}, EMD \cite{yang2009empirical,wu2009diagnosis}, Analytical or Numerical Models \cite{stringer2008gear,stoisser2008comprehensive}, NN \cite{mccormick1997neural,sahraoui2004friction, vijayakumar2006artificial}, Fuzzy Logic \cite{jarrah2004web}, Support Vector Regression (SVR) \cite{omitaomu2006line}, GA \cite{cho2010multivariate, he2001detection}, ARMA \cite{wang2009autoregressive, sinha2002trend} \\
\hline
Pump & Valve impact, score, fracture, piston slap, defective bearing and revolving crank, hydraulic problem & Pump's dynamic responses, generated by a wide range of possible impulsive sources, are very complex; nonlinear, time-varying behavior & Vibration, pressure, acoustic emission & Vibration characteristic frequency, pressure time domain statistical characteristics, sharp pulses and rate of development of stress-waves propagation & FT \cite{ha2002leakage}, STFT \cite{hodkiewicz2004identification, du2007condition, terao2004time}, WT \cite{li2009fault}, Envelop Analysis \cite{jiang2007wavelet}, NN \cite{liang1988prognostics,gibiec2005prediction,engin2007prediction}, Fuzzy Logic \cite{sozen2004performance, perovic2001fuzzy}, Neuro-Fuzzy Hybrid Model \cite{esen2008modelling}, Rough Set \cite{li2006rmine}, PCA \cite{chen2009fault} \\
\end{tabular}
}
\end{table}

\begin{table}[htbp!] %
\centering
{\tiny
\caption*{Table \label{tab:PHMtools} (continued).}
\begin{tabular}{L{1.2cm} L{2cm} L{2cm} L{1cm} L{2cm} L{2cm}}
\hline
\hline
Item & Issue \& failure & Characteristic & Common measures & Common features & Used models \\
\hline
Alternator & Stator faults, rotor electrical faults, rotor mechanical faults & Currents and voltages are preferred for noninvasive and economical testing & Stator currents and voltages, magnetic fields and frame vibrations & Specific harmonic components, sideband components & FT \cite{sottile2006condition}, WT \cite{wan2004vibration, wang2007robust, zanardelli2005failure}, Instantaneous Power Fourier Transform \cite{liu2004online}, Bispectrum \cite{wang2002bispectrum,chow1995three}, High Resolution Spectral Analysis \cite{benbouzid1999induction, hou2003method}, Expert Systems \cite{schoen1995unsupervised, bankert1995model}, NN \cite{penman1994feasibility, filippetti1995neural, byington2004dynamic}, HMM \cite{wu2009integrated}, Fuzzy Logic \cite{sepe1999intelligent, vas1999artificial, filippetti2000recent}, GA \cite{filippetti2000recent}, Higher Order Statistics \cite{arthur2000induction}, Park's Current Vector Pattern \cite{nejjari2000monitoring}, Petri Net \cite{yang2004case}, Kalman Filter \cite{yang2002experiment} \\
\hline

\end{tabular}
}
\end{table}
In addition, there are papers that discuss the prognostics on various components/system including batteries \cite{chan2000available, blanke2005impedance, bhangu2005nonlinear, saha2007integrated, goebel2008prognostics, saha2009prognostics, liu2012lithium, liu2014lithium} (prognostics for batteries appears to be at a more advanced stage than prognostics for structures \cite{schwabacher2005survey}), DC-motor \cite{yang2002experiment, rigatos2009particle, howell2015dc}, electric motors \cite{bonnett2000root, ayaz2006study, ayaz2009fault} boiler tube \cite{majidian2007comparison}, bearing \cite{blair2001diagnosis, ocak2007online, medjaher2012remaining}, engine \cite{yazici1999adaptive}, diesel engine \cite{jardine1989proportional}, wind turbine \cite{guo2009reliability}, gearbox \cite{garga2001hybrid, wang2009robust}, oil \cite{valivs2015contribution}, hydraulic system \cite{liu2015performance}, pumps \cite{goode2000plant, he2012prognostic}, (automatic) transmission \cite{baxter1988proportional, ompusunggu2016kalman} and alternator \cite{siegel2009evaluation}.

In addition, there exist considerable literature on the application of prognostics to aerospace engineering systems with a few of them listed here due to their importance for further study \cite{roemer2001development, skormin2002data, kiddy2003remaining, byington2004dynamic, byington2004data, dong2007hidden, saxena2008damage, bryg2008combining, mazzuchi2008paired, wang2009probabilistic, si2011remaining, lamoureux2012approach, sankararaman2013uncertainty}



\section{Automobile Applications} \label{sec:EA}
In this section, we discuss the details of a few relevant applications of prognostics. There are a few papers that worth attention. For instance, Ref~\cite{rymarz2015reliability} determines a reliability index for the most failure parts and complex systems of two brands of city buses for the period of time failures. The analysis covered damages of the following systems: engine, electrical system, pneumatic system, brake system, driving system, central heating and air-conditioning and doors. Furthermore, the reliability was analyzed based on a Weibull model. Ref~\cite{bouvard2011condition} discusses the maintenance planning for a commercial heavy vehicle. Reference \cite{jardine2001optimizing} discusses work completed to improve the existing oil analysis condition monitoring program being undertaken for wheel motors. Oil analysis results from a fleet of 55 haul truck wheel motors were analyzed along with their respective failures and repairs over a nine-year period.
\subsection{Peer-to-Peer Collaborative Vehicle Health Management}
This section reviews an advanced vehicle diagnostics and prognostics (D\&P) technology initiated by the General Motors company \cite{zhangpeer}. The proposed framework which is called Collaborative Vehicle Health Management (CVHM) is developed to automatically optimize the D\&P algorithms on a host vehicle, using the field data collected from peer vehicles encountered on the road. The objective is to improve the D\&P performance without incurring costs of human intervention. The experimental results on battery RUL prediction show the effectiveness of the proposed framework. This proposed framework has been implemented in a small test fleet as a proof-of-concept prototype.

It is known that the failure modes of vehicles are diverse and vary from vehicle to vehicle. As a result, it is very challenging to achieve accurate and robust D\&P performance for vehicle systems in the field. The traditional approach to D\&P is achieved by introduction of individual faults on bench tests, test vehicles or through accelerated ageing tests and by collecting a large amount of data. This requires a significant amount of algorithm tuning to be done by the development engineers. The proposed CVHM framework is a response to the above challenge, where filed data from peer vehicles are aggregated to automatically optimize the D\&P algorithms for the host vehicle. This is an extension of the decade-long evolving research and development in the area of remote vehicle diagnostics \cite{millstein2002vrm, kuschel2004presenting, carr2005practical, you2005overview, zhang2009connected, byttner2009networked}, which recently has been accelerated due to the advances made in the wireless communication technology and the need for connected vehicle prognostics. The necessary ingredients of a CVHM include
\begin{enumerate}[1)]
\item An onboard CVHM architecture that facilitates efficient aggregation of peer vehicle data, and host vehicle D\&P algorithm adaptation.
\item Intelligent data modelling and statistical decision making technologies that allow the extraction of fault signature, failure precursor, trending information, and other kinds of knowledge that enhances the performance of D\&P.
\item A heterogeneous wireless communication solution that combines cellular network, and opportunistic vehicle-to-vehicle (V2V) communication to allow the exchange of large-volume data between vehicles in a cost-effective way.
\end{enumerate}

\begin{figure}[htbp!]
\centering
\includegraphics[width=3.0in]{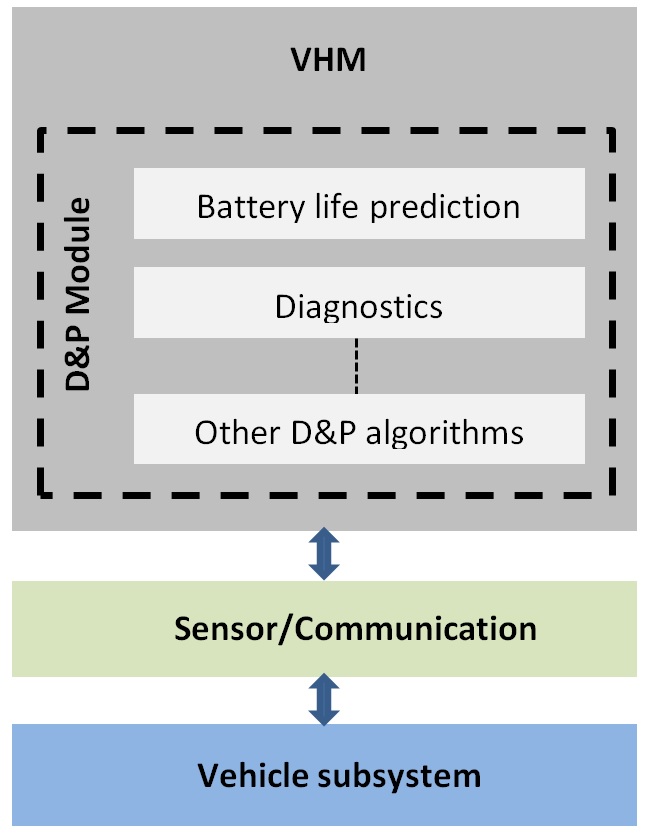}
\caption{A typical VHM system architecture in the state-of-the-art \cite{zhangpeer}.}
\label{fig:TypicalVHM}
\end{figure}

The work in \cite{zhangpeer} addresses the first two items above, using battery RUL for demonstration. A typical vehicle health management system architecture should address the three main tasks associated with CBM, i.e., data collection, feature extraction and decision making. Figure \ref{fig:TypicalVHM} (taken from \cite{zhangpeer}) illustrates a typical architecture. Sensor information regarding particular vehicle subsystem is either directly collected by the VHM ECU that runs D\&P algorithms or is transferred from other ECUs through an in-vehicle communication network. Note that, in real implementations, the VHM ECU may be implemented as a functional module within an ECU, such as a body control module (BCM), that executes control functions. The D\&P module has various D\&P algorithms for different targeted vehicle components or subsystems, such as battery, electrical power generation and storage (EPGS) system, fuel delivery system, etc. The D\&P module processes the sensor information, and generates D\&P results, including the detected anomalies, isolated faulty components, and the predicted RUL of related components. The D\&P algorithms are usually developed, calibrated, and tested through a sophisticated vehicle development process. Once the vehicle is released for production, the D\&P algorithms and the associated calibration values are usually fixed. If major updates on the onboard algorithms are needed, an ECU reprogramming can be done after the vehicle is usually called to a dealer service shop. Lately, the technology of remote ECU refresh is maturing, which may allow the ECU reprogramming to be done remotely through telematics connections.

\begin{figure}[htbp!]
\centering
\includegraphics[width=4.0in]{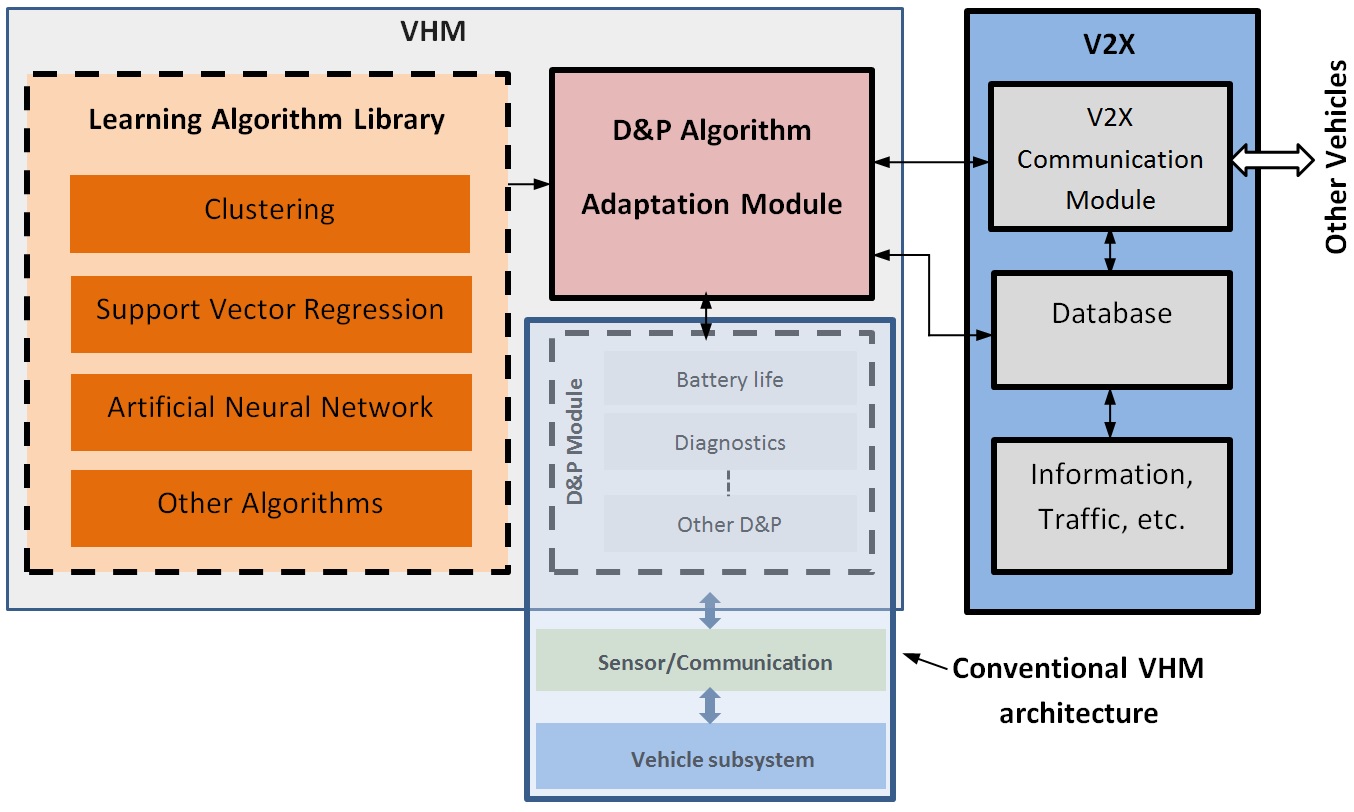}
\caption{CVHM system architecture proposed in \cite{zhangpeer}.}
\label{fig:ProposedVHM}
\end{figure}

The CVHM system architecture proposed in \cite{zhangpeer} should address the tree main ingredients of a realizable CBM system. Figure \ref{fig:ProposedVHM} demonstrates the proposed CVHM system architecture.
The newly added V2X ECU provides the wireless communication interface in order to exchange vehicle health related data between the host vehicle and peer vehicles. V2X represents vehicle-to-vehicle or vehicle to infrastructure. The V2X ECU stores the data in an onboard database. The VHM ECU has an algorithm adaptation module and a learning algorithm library, in addition to the regular D\&P module. The algorithm adaptation module makes use of appropriate learning algorithms to process the vehicle health related data stored in the onboard database in order to tune and optimize the calibration values within the D\&P module.

The advantage of CVHM can be understood based on the following example. A battery life prediction algorithm usually implements an ageing model that specifies how the battery internal resistance grows given the number of charge-discharge cycles. There are parameters in the ageing model that specifies the growth rate of the battery internal resistance, which is critical in battery life prediction. These parameters are typically calibrated using accelerated ageing test during the vehicle development process, and applied to across the board to all vehicles. However, it is difficult for a pre-calibrated model to account for the intrinsic diversity of usage patterns and environment impacts. The fact is that batteries for the same battery/vehicle model may have different life span that ranges from 1 year to 10+ years. At the same time, with large enough vehicle population, for any given vehicle, chance is high that there are peer vehicles with similar usage profiles that have been used for longer time, and therefore have gone further ahead in the ageing process. With CVHM, field data from these peer vehicles can be used to fine tune the growth rate in the battery ageing model, and consequently achieve higher prediction performance.

The general framework to develop model-based prognostics for RUL prediction involves the following steps. First, one or more fault signatures are identified to characterize target system’s state of health (SOH), $Z=f(SOH)$. Depending on applications, these fault signatures may be assessed either directly or indirectly. For example, in the application of Starting, Light, Ignition (SLI) battery life prediction, multiple fault signatures have been proposed. The second step is to establish the failure criteria for fault signatures with respect to specific applications. That is, if $Z>Z_0$, a system failure is declared, where $Z_0$ is a threshold. For example, one of the main functions for SLI battery is to crank the engine. As battery ages, its SOH deteriorates, and so does its cranking capability. One of the fault signatures, cranking resistance, increases during the ageing process. When the cranking resistance reaches certain level, the engine can hardly be started. This is when a battery failure is declared. The failure criteria are highly application specific, and usually require careful calibration. The third step is to establish a system-ageing model that specifies how the fault signatures evolve with respect to usage. That is,
\begin{equation*}
Z = Z(L;\theta),
\end{equation*}
where $L$ is a set of variables that characterize the usage profile of the target system, and $\theta$ is a set of parameters that specify the detailed relationship between the usage and the fault signature evolution.

Extensive previous research has been conducted, and multiple SLI battery fault signatures have been identified, including minimum cranking voltage, delta V, cranking power, voltage residual, and cranking resistance. For instance, the cranking resistance increases in an accelerated ageing experiment. In \cite{zhangpeer}, a few static parametric models are adopted, including polynomial curve fitting. The algorithm development is based on a 3rd order polynomial model due to its structural simplicity. Each fault signature is modeled by the following equation
\begin{equation*}
\hat y(t) = p_1 t^3 + p_2 t^2 + p_3 t +p_4,
\end{equation*}
where $\hat y$ is predicted fault signature value, $t$ is the battery age in terms of service time, and $p_1,p_2,p_3$ and $p_4$ are model parameters. Since both SOC and battery temperature can affect battery fault signature, different models have to be learned for different SOC and temperatures. The battery RUL is defined as
\begin{equation*}
RUL =  arg \min_t [\hat y(t)=y_0] - t_{current},
\end{equation*}
where $y_0$ is a predefined threshold, and $t_{current}$ is the current battery age. The ageing model calibrated with accelerated ageing test may not be able to characterize the ageing process in the field. In the proposed CVHM, the ageing model is adapted using the data from peer vehicles that have gone further in the ageing process. Let $y_H(t_j)$ be the fault signature value measured or estimated by the host vehicle at time instant $t_i$, where $j = 1 \cdots J$ and $J$ is the current time index for the host vehicle. Let $P_{H,1}, P_{H,2}, P_{H,3}$, and $P_{H,4}$ be the ageing model parameters maintained by host vehicle, and  $P_{P_k,1}, P_{P_k,2}, P_{P_k,3}$, and $P_{P_k,4}$ be the ageing model parameters used by peer vehicle $P_k$, where $k = 1 \cdots K$ and $K$ is the number of peer vehicles. The model adaptation procedure is as follows,
\begin{enumerate}[1)]
\item Estimate host vehicle fault signature values using peer vehicle's ageing model parameters, which yields,
    \begin{equation*}
    \hat{y}_{H,P_k}(t_j) = p_{P_k,1} t_j^3 + p_{P_k,2} t_j^2 + p_{P_k,3} t_j + p_{P_k,4},     
    \end{equation*}
    where $\hat{y}_{H,P_k}(t_j)$ indicates the estimate of host vehicle fault signature using the ageing model from peer vehicle $P_k$.
\item Calculate the corresponding estimation error for the ageing model from each peer vehicle $P_k$ as,
    \begin{equation*}
    R_{H,P_k} = \sum_{j=1}^J [\hat{y}_{H,P_k}(t_j)- y_H(t_j)]^2.
    \end{equation*}
\item Pick N models with the smallest error. Without loss of generality, the corresponding peer vehicles can be represented as $P_{k_1}$, $P_{k_2}$, $\cdots$, $P_{k_N}$. In the experiment presented in this paper, $N$ is set to 3.
\item Calculate the adjusted host vehicle fault signature values, $\bar{y}_H(t_j)$, by averaging the fault signature values based on the selected peer vehicle's ageing models,
    \begin{equation*}
    \bar{y}_H(t_j) = \frac{1}{N} \sum_{n=1}^N \hat{y}_{H,P_{k_n}}(t_j).
    \end{equation*}
\item Update the host vehicle ageing model, using the adjusted fault signature values
     \begin{equation*}
    {p_{H,1},\cdots, p_{H,4}} = arg \min_{p_1,\cdots p_4} \sum_{j=1}^J [\hat{y}_{H}(t_j)-\hat{y}(t_j)]^2,
    \end{equation*}
    where $\hat{y}(t_j) = p_1 t_j^3 + p_2 t_j^2 +p_3 t_j +p_4$.
\end{enumerate}
The adjusted ageing model parameters $p_{H,1},\cdots, p_{H,4}$ are used for future battery RUL prediction of the host vehicle.
\subsubsection{System Implementation}
The CVHM architecture proposed in \cite{zhangpeer} has been implemented in a three-vehicle test fleet for the battery RUL prognosis application. To reduce the development cycle and cost, the test fleet is constructed in a way that one host vehicle implements the full CVHM architecture, and two peer vehicles implement only the V2X module. Each of the two peer vehicles maintain a database of battery D\&P data from multiple batteries, which simulates the situation where data from multiple peer vehicles can be transferred to the host vehicle for D\&P algorithm adaptation.

For the host vehicle prototype implementation, there are three major hardware components. The first one is a dSpace® MicroAutoBox (MAB) that has direct connection with the sensors on the battery. It employs the functions of data acquisition, signal pre-processing, and fault signature generation. During each vehicle cranking process, the MAB generates multiple battery-status related parameters, including battery temperature, SOC, cranking resistance, minimum cranking voltage, cranking powering, delta V, voltage residual. The third major hardware component is a V2X communication laptop (an HP® Compaq 6910P with the OS of Linux Ubuntu 10) that communicates with the VHM laptop through TCP/IP based connection. The V2X laptop implements the V2X module that interacts with peer vehicles and infrastructure through a wireless communication to exchange data. It maintains a MySQL® database server to organize the data as well as manage the retrieval requests from the VHM module. The V2X laptop also serves as the driver interface module to provide battery health information to the end user.

\subsubsection{Experimental Results}
The CVHM system has been validated using the $JBI_Aging_2008$ data set. In this data collection effort, 15 batteries from different suppliers were aged from fresh to the end of life through an accelerated ageing process. The battery age varies from 8 to 16 weeks. During the ageing process, weekly cranking tests were conducted on a test vehicle for each battery after it was conditioned to 100\% state of charge (SOC) and the temperature of 25°C. Battery current, battery voltage, and engine RPM were collected during cranking. After data cleaning, there are totally 1710 cranking data files that have adequate data for 14 batteries. Among these fault signatures, cranking resistance appears to be better SOH indicators than others, due to its consistency and monotonic correlation with the battery age. Therefore, we selected the cranking resistance as the fault signature in the rest of the experiments. The accuracy specifies the difference between predicted value and the actual value. The precision specifies the spread of the predicted values.
\subsubsection{Simulation Results}
\begin{figure}[htbp!]
\centering
\includegraphics[width=4.0in]{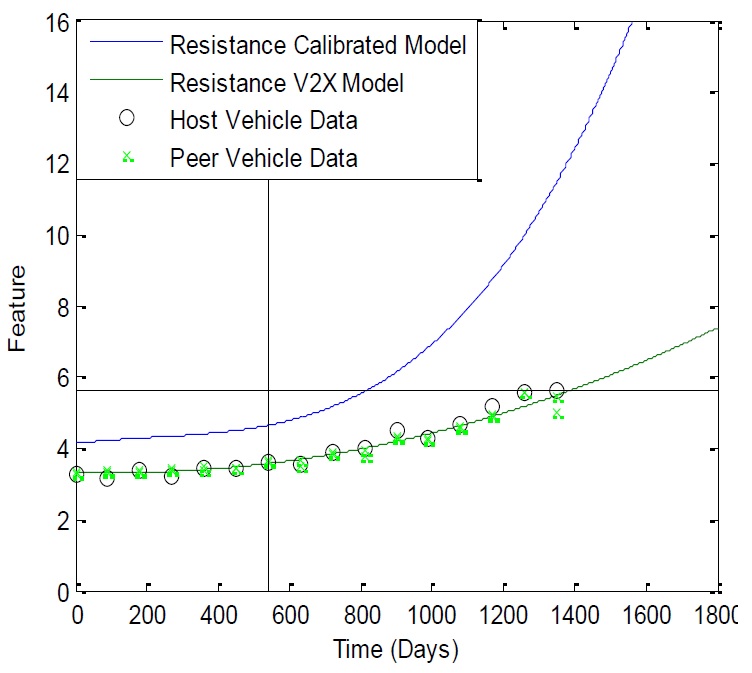}
\caption{Comparison of battery RUL prediction with pre-calibrated model and adaptive model \cite{zhangpeer}.}
\label{fig:BatteryCVHM}
\end{figure}

Figure \ref{fig:BatteryCVHM} illustrates the battery RUL prediction results during one particular ignition cycle. At this particular ignition cycle, the host vehicle battery has been in service for 540 days, assuming each week of accelerated ageing corresponding to about 90 days of real-world driving. The cranking resistance has increased from the initial value, but is still significantly lower than the end of life threshold indicated by the black horizontal line. The initially calibrated ageing model, as shown by the blue line, predicts the RUL is about 250 days, since the cranking resistance is predicted to pass the threshold in about 250 days. This prediction is very different from the actual cranking resistance data that are shown by the black cycles. At the same time, the host vehicle has access to the data from peer vehicles’ batteries, of which the data from nearest neighbors are shown by the green crosses. Following the model adjustment procedure presented before, an updated battery ageing model is obtained, and shown by the green line. The updated ageing model traces the actual cranking resistance very well, and provides a fairly accurate RUL prediction.
\subsubsection{Preliminary penetration analysis}
A discussion is presented on how many peer vehicles are needed to achieve specific RUL prediction performance. The performance of RUL prediction can be measured by the accuracy and the precision. The relations indicate that under the i.i.d. assumption, the RUL estimation will have zero expected error, which is very desirable. And the error spread of the CVHM-based prediction is reduced by a factor of $\frac{\sqrt{n+1}}{n}$ from the single vehicle battery RUL prediction, which shows why the CVHM framework enhances the prediction performance.
\subsection{Automatic transmission: Wet-clutch system}
Nowadays, automatic transmissions have become a popular choice in commercial vehicles and been widely used in off-road/heavy-duty vehicles. As is obvious from its name, an automatic transmission is a transmission that shifts power or speed by itself. The key element that enables automatic powershifting
or speed-selection in automatic transmissions is a wet friction clutch. The power transmission from the engine to the wheels through wet friction clutch is based on the friction occurring in lubricated contacting surfaces. A wet friction clutch (hereafter called wet clutch) is lubricated by an automatic transmission fluid (ATF) having a function as a cooling lubricant cleaning the contacting surfaces and giving smoother performance and longer life. For high-power applications, the clutch is typically assembled with multiple friction and separator discs. The friction disc is made of a steel-core-disc with friction material bonded on both sides and the separator disc is made of plain steel. An electromechanical hydraulic actuator is usually used for engaging/disengaging a wet clutch. This actuator consists of some main components, such as a piston, a return spring which is always under compression and a hydraulic group consisting of a control valve, an oil pump, etc. To engage the clutch, pressurized ATF is controlled by the valve to generate a force acting on the piston.

A number of researchers have explored and developed model-based prognostics techniques for wet clutches. Yang et al. (see \cite{yang1997theoretical, yang1998theoretical}) developed a physics-based prognostics model by considering that the degradation occurring in a wet clutch is due to thermal effect alone in the friction materials. To this end, a dedicated invasive and destructive test, i.e., thermal gravimetric analysis, is required for identifying some parameters for the prognostics model. Since the degradation mechanism occurring in the clutch friction material is not only due to thermal effect but also another major mechanism namely adhesive wear (see \cite{gao2002friction, gao2002microcontact, ompusunggu2015distinguishing}), the assumption made within the prognostics method in \cite{yang1997theoretical, yang1998theoretical} is, therefore, too oversimplified. Moreover, this approach would be difficult to implement by the end users when the complete design data of a wet clutch system are not available. Prognostics algorithms for the ATF (i.e., lubricant) of wet clutches have been also developed and reported in the literature. References \cite{calcut2004estimating} and \cite{sarkar2005real} developed an empirical degradation model for predicting the lifetime of ATF based on an SAE\#2 modified plate test, in which the energy per shift and bulk lubricant temperature are used as input parameters. The degradation model applies only to a specific ATF under certain operating conditions. Furthermore, Ref \cite{hirthe2005monitoring} developed a prognostics methodology using extra lubricant sensor which is immersed in ATF. The sensor provides an electrical signal indicating in real time the chemical condition of the lubricant to be monitored. Three parameters, namely 1) total acid number (TAN), 2) delta oxidation (OX), and 3) HPDSC induction time (MIN), can be derived from the sensor readings. An empirical model was developed to predict the RUL of ATF based on the three parameters. Because of its robustness, hybrid prognostics approach under the Bayesian framework (e.g., Kalman filtering) has been attracting a number of researchers nowadays and been successfully applied to various applications like bearings, batteries, material crack growth, electrolytic capacitors, etc.

Demands of low-cost prognostics tool for automatic transmission clutches (i.e., based on measurement data from sensors typically available) by industry have increased since the last few years. In \cite{ompusunggu2016kalman}, a prognostics tool is developed by fusing a newly developed degradation model with the measurable pre-lockup feature under the extended Kalman filtering framework. As this feature can be extracted from sensory data typically available in wet clutch applications, the developed prognostics tool, hence, does not require extra cost for any additional sensor. New history data of commercially available wet clutches obtained from accelerated life tests using a fully instrumented SAE\#2 test setup have been acquired and processed. The experimental results show that the prognostics algorithm developed outperforms the early developed prognostics algorithm, which is based on the weighted mean
slope method (i.e., data-driven approach). It is shown that the clutch remaining useful life estimations with the novel prognostics algorithm remain in the desired accuracy region of 20\% with relatively small uncertainty interval in comparison .with the early developed prognostics algorithm. In this framework, empirical or physics-based degradation models are fused with measurement data (i.e., feature) in order to improve the RUL estimation.

\subsection{Alternator}
A vehicle alternator shares many similarities to AC generators and induction motors, and studies have shown that bearing failure is responsible for 40\% of the failures for induction motors, making it the most common mechanical failure \cite{onel2006detection}. The alternator is coupled to the engine by either a v-belt or a serpentine belt pulley system and a higher than normal level of belt tension can provide greater lateral loads on the bearing and reduce its life. There is also a great deal of research in bearing health monitoring and prognostics, which is likely not only due to bearings being an important component for rotating machinery, but also due to the bearing geometry, there are specific fault frequencies that are seen in the vibration spectrum \cite{qiu2003robust}. 

A vehicle alternator's overall function is to charge the vehicle battery as well as power the electrical auxiliaries; an alternator that is degraded or failed will ultimately result in the inability to use these additional auxiliaries and an increased potential for a dead battery and stalled car. Considering the importance of the alternator in the overall functioning of the vehicle and perhaps in particular for military vehicles for which mission success is dependent on the use of surveillance equipment or other features that require a properly functioning vehicle electrical system; knowledge of the health status of the alternator is useful information that can support logistical, tactical and maintenance planning efforts. 

The vehicle alternator is essentially a rotating machine that generates a 3-phase alternating current that is rectified by a set of diodes in order to produce a DC current with low ripple content; the vehicle alternator shares many similarities with a generator, and to some degree electrical motors, and in turn some of the common health monitoring and prognostic techniques, as well as common failure modes, for motors and generators are applicable. Considering the similarity between vehicle alternators, electric motors and generators, a general methodology for assessing the health of rotating electro-mechanical components was developed and demonstrated for an automotive alternator component. 


The approach applies domain specific knowledge, along with processing the data and extracting features from the time domain signal, as well as the order spectrum, to train machine learning algorithms, such as statistical pattern recognition, logistic regression, or a self-organizing map to assess the health of the
vehicle alternator.

For this particular case study, three components of the alternator were monitored, the alternator bearings, stator windings and diodes, with respective features from the electrical or vibration signals that are correlated to the degradation of each of those components. The processing of the alternator tachometer signal and the vibration and electrical signals into the order spectrum provides a way to extract relevant information that is indicative of bearing, diode and stator health. 

Three health assessment algorithms were highlighted for this particular study with both the logistic
regression method and the self-organizing map method performing quite well with the logistic regression technique having a type I and II error of 5\% \cite{siegel2009evaluation}. The overall framework utilized in this case could be extended to other applications, and assessing the component health over time is a pre-requisite for prognostics in which further work could look at developing a remaining useful life prediction technique for vehicle alternators.

\section{Challenges And Opportunities}
This section discusses some of the challenges and difficulties associated with developing prognostics models. Attention is given to those aspects that need to be further investigated for reliable methods to be used in real-life situations. It is essential to develop methods that could utilize the available data and accurately incorporate them into the models. On the other hand, the operating conditions of machines in the real-life is different from the experimental test done in the lab. The final consequent of these operational complexities can greatly diminish the accuracy of the prognosis output. 

In most of the literature, it is only tried to predict a specific failure mode of one individual component without considering the interaction of the other component with the asset under study or with the operating environment. It is important to look at particular areas in which new ideas and improvements could offer opportunities for enhanced prognostics. These include: proper incorporation of CM data into reliability; proper incorporation of incomplete trending data; how the maintenance actions and operating conditions effect the results; what would be the best non-linear or linear model to describe the actual degradation vs the predicted one; how the failure interactions should be considered; how to verify the accuracy of the assumptions. These challenges are discussed more in \cite{heng2009rotating}.

%% file: Conclusion.tex
\section{Concluding remarks} \label{CR}
In this paper, a review of the state-of-the-art models, methods and algorithms to engineering prognostics is presented with a focus on practical aspects. The advantages and weaknesses of the methods and models and briefly reviewed which is later revisited within the context of prognostics and health management tool development. A separate section of the work is devoted to the prominent instances of applications of prognostics to components that are highly used in engineering system and a few automobile-related examples are discussed in details.   
\begin{itemize}
\item The remaining useful life estimation models are categorized into 1) data-driven 2) model-based approaches 3) hybrid approaches
\item The selection of the best model depends on the level of accuracy and availability of data.
\item In cases of quick estimations which are less accurate, the data driven method is preferred, while the physics-based approach is applied when the accuracy of estimation is important.
\item For most industry applications, physics-based models might not be the most practical solution since fault type is question is often unique from component to component and is hard to be identified without interrupting operation. However, physics-based models may be the most suitable approach for cost-justified applications in which accuracy outweighs other factors and physics model remain consistent across systems, such as in air vehicles. They also generally require less data than data-driven models.
\item Data-driven models may often be the more available solution in many practical cases in which it is easier to gather data than to build accurate system physics models.
\item The modelling approach selected must be fit for purpose. All models are subject to underlying assumptions of implementation constraints that restrict their applicability to certain types of problems. Common issues relate to the following: (1) amount,type and quality of data required;(2) effect of process and measurement noise on data; (3) type of repair (perfect,imperfect);(4) number of failure modes that can be simultaneously or collectively modelled; and(5) whether or not novel failure types can be managed.
\item Although all models require some understanding of the underlying failure process, some approaches require detailed technical knowledge of failure mechanisms and data collection processes for model implementation. Of the models discussed, Neural Networks require minimal understanding about the processes governing failure to apply, while expert and fuzzy systems require a medium amount;physical models require comprehensive knowledge pertaining to all physical mechanisms and environmental factors influencing equipment failure for their successful application.
\item All models require data for design, parameter definition and validation. The completeness requirement for data sets varies between models. Organisations need to improve data quality management processes if they wish utilize prognostic modelling more widely to support asset decision making.
\item Data requirements for diagnostics are often different to data required for prognostic modelling.
\item Not all models supply confidence limits on their predictions,which is necessary to manage the uncertainty in setting priorities, decision making and for practical risk management. Most Neural Networ kmodels in particular cannot determine confidence bounds for an estimate.
\item Not all models are able to predict RUL with the same level of accuracy and precision. Therefore business requirements need to be clearly articulated and incorporated into model selection processes.
\item Few of the current approaches are suitable for imperfect repair. Exceptions include Duane growth(type of aggregate reliability function), proportional intensity model(variant of PHM)and Hidden/Semi-hidden Markov Bayesian models.
\item Most models are better suited for component failure mode RUL estimation (i.e., one dominant failure mode) yet are often applied at a system level due to data availability on time constraints. Not surprisingly, results on system failure prognostics are mixed. Successful applications of system level RUL estimation are mainly for equipment where one failure mode(or at most a very small number of failure modes) dominates overall system reliability. More examples illustrating the effect of overlapping failure mechanisms are required to verify the suitability of techniques for system-level prognostics.
\item Few case studies are available that illustrate the application of the prognostic models published in the academic literature to real world problems in realistic operating environments (i.e., systems with overlapping failure modes, subject to common mode failures or undergoing highly variable process changes). This is an area where significantly more work needs to be published to verify that prognostic models are useful for main stream asset management decision making of routine assets.
\item Mathematical or computing complexity currently limits current use of many approaches to industry practitioners. Although commercially available software tools eliminate the need for software programmers, developing a meaningful model is often more involved than the suppliers’literature implies. Where software is not available, users should not underestimate the resources required to develop the code to instantiate a model. More information provided by authors of journal articles about the level of modelling skill and time expended when building the particular model presented in a published work would be helpful to industry practitioners to identify resource requirements.
\item Appropriate model selection for successful practical implementation, requires both a mathematical understanding of each model type, and also an appreciation of how a particular business intends to utilize the models and their outputs.
\item The ability of PHM to estimate RUL for prognosis of varied faults in complex systems is uncertain and likely to be an ongoing challenge.

\end{itemize}